\documentclass[conference]{IEEEtran}
\IEEEoverridecommandlockouts
\usepackage{cite}
\usepackage[dvipsnames]{xcolor}
\usepackage{amsmath,amssymb,amsfonts}
\usepackage{algorithm,algorithmicx}
\usepackage{algpseudocode}
\usepackage{graphicx}
\usepackage{textcomp}
\usepackage{rtsched}
\usepackage{todonotes}
\usepackage{multicol}
\usepackage[utf8]{inputenc}
\usepackage[english]{babel}
\usepackage{amsthm}
\usepackage{soul}
\usepackage{comment}
\usepackage{balance}
\usepackage{xcolor}

\usepackage{xkcdcolors} 
\usepackage{tikz}
\usepackage{wrapfig}
\usetikzlibrary{backgrounds,decorations.pathreplacing}
\usetikzlibrary{arrows,shapes}
\usetikzlibrary{tikzmark}
\usetikzlibrary{calc}

\usepackage[edges]{forest}
\usetikzlibrary{arrows.meta}
\colorlet{linecol}{black!75}

\usepackage{algpseudocode}
\algrenewcommand\textproc{}

\newcommand{\highlight}[2]{\colorbox{#1!17}{#2}}

\usepackage{xspace}
\newcommand{\paranoid}{{\sc \textit{Paranoid}}\xspace} 

\newcommand{\win}{\Omega}

\begin{document}


\title{SchedGuard: Protecting against Schedule Leaks Using Linux Containers
}

\author{
  \IEEEauthorblockN{
    Jiyang Chen\IEEEauthorrefmark{1},
    Tomasz Kloda\IEEEauthorrefmark{2},
    Ayoosh Bansal\IEEEauthorrefmark{1},
    Rohan Tabish\IEEEauthorrefmark{1},
    Chien-Ying Chen\IEEEauthorrefmark{1}\IEEEauthorrefmark{3},
    Bo Liu\IEEEauthorrefmark{1}\IEEEauthorrefmark{3},\\
    Sibin Mohan\IEEEauthorrefmark{1},
    Marco Caccamo\IEEEauthorrefmark{2} and
    Lui Sha\IEEEauthorrefmark{1}\\
    \thanks{\IEEEauthorrefmark{3} Chien-Ying Chen and Bo Liu are now with {NVIDIA} Corporation, USA}
  }
  \IEEEauthorblockA{\IEEEauthorrefmark{1}\textit{University of Illinois at Urbana-Champaign}, USA,~~~~\IEEEauthorrefmark{2}\textit{Technical University of Munich}, Germany\\ \{jchen185, ayooshb2, rtabish, cchen140, boliu1, sibin, lrs\}@illinois.edu,~\{tomasz.kloda, mcaccamo\}@tum.de
  }
}

\maketitle

\begin{abstract}

Real-time systems have recently been shown to be vulnerable to timing inference attacks, mainly due to their predictable behavioral patterns. 
Existing solutions such as schedule randomization lack the ability to protect against such attacks, often limited by the system's real-time nature.
This paper presents ``\textit{SchedGuard}'': a temporal protection framework for Linux-based hard real-time systems that protects against posterior scheduler side-channel attacks by preventing untrusted tasks from executing during specific time segments.
SchedGuard is integrated into the Linux kernel using cgroups, making it amenable to use with container frameworks.
We demonstrate the effectiveness of our system using a realistic radio-controlled rover platform and synthetically generated workloads.
Not only is SchedGuard able to protect against the attacks mentioned above, but it also ensures that the real-time tasks/containers meet their temporal requirements.

\end{abstract}

\begin{IEEEkeywords}
Real-Time, CPS, Response time analysis, Linux Containers, Security 
\end{IEEEkeywords}

\begin{tikzpicture}[remember picture,overlay]
  \node[anchor=south west,inner sep=0pt] at ($(current page.north east)+(-3.8cm,-4.5cm)$) {
     \includegraphics[scale=0.11]{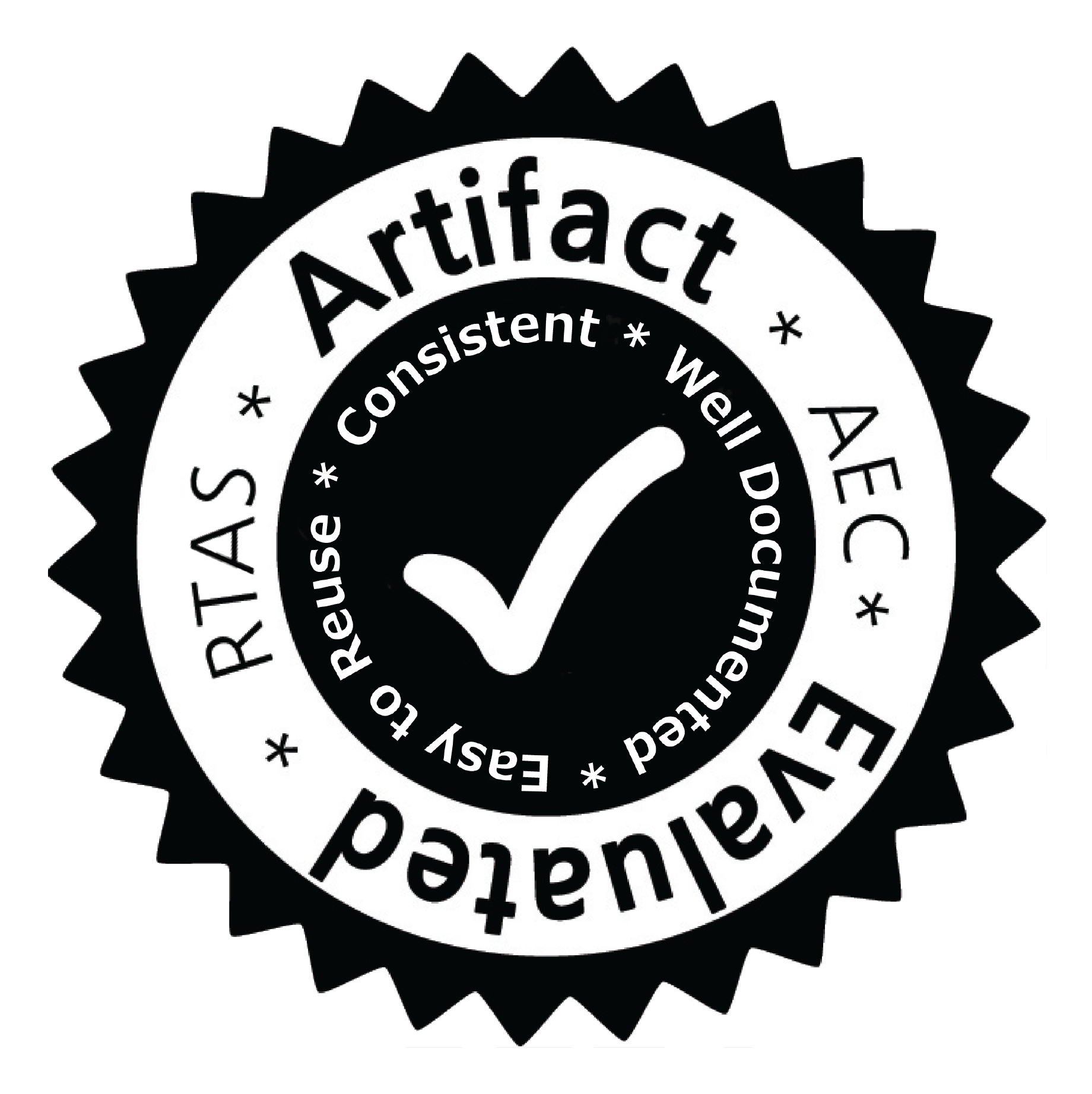}
  };
\end{tikzpicture}


\theoremstyle{definition}
\newtheorem{definition}{Definition}
\newtheorem{theorem}{Theorem}
\newtheorem{corollary}{Corollary}[theorem]
\newtheorem{lemma}{Lemma}

\section{Introduction}

Many legacy real-time systems still run on single-core processors.
Various works in the real-time security community have demonstrated how schedule-based attacks can compromise system security when running along with other trusted and useful tasks in the system~\cite{chen2018scheduleak}. 
Schedule randomization based approaches have been proposed to deter against such attacks~\cite{yoon2016taskshuffler,yoon2019taskshuffler++,chen2018reorder}.
However, recently  Nasri et al.~\cite{nasri2019pitfalls} suggested that randomization-based approaches might fail to defend against schedule-based attacks.


The success of schedule-based attacks relies on how soon the attacker can run relative to the completion of the victim task.
There are various works in the security literature that deploy the above-mentioned exploit for a successful attack~\cite{nasri2019pitfalls}. 
These attacks target the exact duration of when the victim finishes its execution or interacts with the outside world through I/O channels.
Examples of such attacks in the cyber-physical research community include exploits such as bias-injection attacks~\cite{teixeira2012attack}, zero-dynamics attacks\cite{teixeira2012attack,teixeira2012revealing,park2016adversary,kim2016zero,jafarnejadsani2017dual,kim2018zero}, and replay attacks~\cite{mo2009secure}. 
The schedule based attacks exploit have also been deployed in the context of general-purpose computing. In this context, attacks such as cache-timing attacks are common. 
These attacks steal or compromise the victim task's data integrity by scheduling themselves right after the completion of the victim task where important crypto-related information of the victim task might still be available in the shared caches or DRAM. 
Cache-flush based defense mechanism can help defend against such attacks.

For the schedule-based attacks to be effective, they have to be deployed/executed within a certain time window after the completion of the victim task. 
In this paper, we define this time for the attacker task as \textbf{attack effective window} ($AEW$). 
Any time greater than $AEW$ makes the attack ineffective.
An example of $AEW$ has been successfully demonstrated in \emph{ScheduLeak}~\cite{chen2018scheduleak}.
The authors determined the $AEW$ for a control output overwrite attack to be 8.3ms. 

Depending upon the timing relation between when the attacker launches the attack on the victim task, the schedule-based attack has been categorized into different attack  categories~\cite{nasri2019pitfalls}: (a) \textbf{posterior attack model:} an attack is launched after the victim has completed its execution;
(b) \textbf{anterior attack model:} an attack is mounted before the execution of the victim task; (c)~\textbf{pincer attack model:} follows a hybrid approach that aims at combing the posterior and anterior attack approaches where the attacker analyzes the victim task at load time and monitors its behavior after the victim task has completed execution; (d) \textbf{concurrent attack model:} performs the attack while the victim is running and can be mounted by executing between the execution window of the victim task's job. In this paper, we only consider defense against the posterior attack model.

Existing randomization-based defense approaches~\cite{yoon2016taskshuffler,yoon2019taskshuffler++,chen2018reorder} against schedule-based attacks are either very ineffective or  incur large overheads, thus affecting response time \cite{nasri2019pitfalls} of the victim task as well as system schedulability. 
We implement the approach using cgroup, which is one of the main techniques used for enabling Linux containers.
This allows us to take advantage of the resource isolation and protection provided by the container. 
In this work, we propose a new systematic approach called \emph{SchedGuard} (schedule guard) that blocks untrusted tasks from running right after the victim task. The main contribution of this work is: 

\begin{itemize}
    \item We propose a temporal isolation mechanism, \emph{SchedGuard}, to defend against posterior schedule-based attacks targeting cyber-physical systems. 
    \item We analyze and evaluate the system's schedulability under different scheduling mechanisms.
    \item We propose a new security-oriented scheduling policy that prioritizes $AEW$ coverage while maintaining system schedulability and use simulation to evaluate its~\mbox{performance.}
    \item We implement our proposed \emph{SchedGuard} approach in the Linux scheduler and demonstrate its effectiveness on commercial-off-the-shelf (COTS) RC cars. 
\end{itemize}

\section{System and adversary model}

This section discusses the system and threat models.

\subsection{System model}
We consider a uniprocessor platform that runs real-time periodic tasks. 
Each task $\tau_i$ is characterized by \((C_i, T_i, \phi_i)\) where \(C_i\) is its worst-case execution time, \(T_i\) is its period, and \(\phi_i < T_i\) its initial offset and we assume that the deadline is equal to period ($D_i=T_i$).
All the above parameters are positive integers.
The tasks are scheduled using
fixed-priority preemptive scheduling algorithms and every priority is assigned by \emph{Rate Monotonic} (\emph{RM}) algorithm (\emph{i.e.,} the shorter the period, the higher the priority)~\cite{liu1973scheduling}.
Table~\ref{tab:notation_summary} summarizes the task sets notations relative to the task~$\tau_i$'s~priority.  

\begin{table}[tbp]
\caption{Task sets notation.}
\centering
{
\footnotesize
\def\arraystretch{1.3}\tabcolsep=5pt
\begin{tabular}{|c|clp{1.2cm}|}
\hline
Notation     &  \multicolumn{3}{c|}{Description} \\
\hline
$hp(i)$ &  & tasks with higher priority & than $\tau_i$ \\
$lp(i)$ &  & tasks with lower priority  & than $\tau_i$ \\
$thp(i)$ & trusted & tasks with higher priority & than $\tau_i$ \\
$tlp(i)$ & trusted & tasks with lower priority & than $\tau_i$ \\
$uhp(i)$ & untrusted & tasks with higher priority & than $\tau_i$ \\
$ulp(i)$ & untrusted & tasks with lower priority & than $\tau_i$ \\
\hline
\end{tabular}
}
\label{tab:notation_summary}
\end{table}

The worst-case response time $R_i$ for task $\tau_i$ is  the longest time between the release of a job of the task~$\tau_i$ until its completion. The task is schedulable if its worst-case response time is less than or equal to its deadline~($R_i \leq D_i$).

\subsection{Threat model} 

In our threat model, we assume the goal of the attacker is to successfully perform a posterior scheduler-based attack.
We categorize attacker's capabilities into technical ones and operational ones.

Technical capabilities refer to the assumptions about the attacker's knowledge regarding the target platform and victim application. 
We assume that the attacker has access to a copy of the target hardware and software system, including the victim's binary, such that they have full knowledge about the hardware platform as well as the victim task's execution time and periods.
In the case where scheduling related parameters cannot be determined offline (e.g. task initial offset), attacker can obtain such information after getting inside the system and using techniques such as the Scheduleak \cite{chen2018scheduleak}. 
We also assume that the attacker can analyze the application on the target system and use other exploits for remote code execution~attacks. 

Operational capabilities captures the attacker's abilities to implement the attack.
CPSes nowadays usually have a communication module that connects with the outside world using WiFi, radio or cellular. 
These are remote attack surfaces that adversaries can exploit to get inside the target system. 
In this paper we consider only remote attacks in which the adversary does not have physical access to actually deploy the attack but can exploit unsecured wireless network or wireless configurations as explained in the following examples.
Some drones and radio-controlled vehicles allow users to control through tablets and mobile phones. 
However, these communication protocols are considered insecure and can be exploited by attackers to install malware \cite{yaacoub2020security}.
Besides, communication modules use legacy software with unpatched, known vulnerabilities.
For example, the security team in Tencent was able to remotely hack a Tesla through legacy browser software~\cite{nie2017free}.
They noticed that the Tesla web browser used an old version of QtWebkit that has many vulnerabilities. 
Through two exploits they achieved arbitrary code execution in the center display system in the~Tesla. 

We assume a capability-based security system where a program requires certain "capabilities" to achieve certain operation. 
For example, starting with Linux 2.2 superuser privileges are divided into distinct units known as capabilities and they can be independently enabled and disable for each process. 
Only superuser can assign capabilities to other processes.
\(CAP\_SYS\_NICE\) is a Linux capability that gives a process permission to change parameters such as scheduling policy, period and priority. 
\(CAP\_SYS\_RAWIO\) is a capability to allow a process to write to an I/O device.
To write to a device I/O one needs to acquire this capability. 
We argue that it is not uncommon for communication modules (such as radio) to have access to hardware I/O but it is unnecessary for them to have the capability to change scheduling parameters.
The attacker could gain device I/O access by remotely exploiting the communication module but cannot gain the capability to modify scheduling system. 
Note that although we assume the attacker can access I/O, we do not consider Denial-of-Service attack on I/O in this paper. 
The DoS attack can be mitigated, for example, by rate limiting some system critical resources~\cite{chen2019container}.

Inside the target system, we assume that the attacker does not have the ability to exploit kernel vulnerabilities and gain root privilege.
Although in the aforementioned attack, the security team was able to achieve privilege escalation to gain root access in the system, they attribute their success to the fact that the system was using an old version of the Linux kernel (2.6.36) which does not have many exploit mitigation applied.
They also commended Tesla's response that patched all famous kernel vulnerabilities in the old kernel and also introduced new kernel (4.4.35) in newer models. 
With security concerns rising for CPS, it will become difficult for attackers to achieve privilege escalation in newer generation of CPS.
However, these do not stop the attacker from getting into the system and launch attacks that do not require root~privileges.

We formally define the attack effective window~($AEW$). 
\theoremstyle{definition}
\begin{definition}
Attack effective window~$\win > 0$ is the time period during  which  scheduled-based  attacks are effective and ineffective otherwise.
\end{definition}

\begin{figure}[hbt]
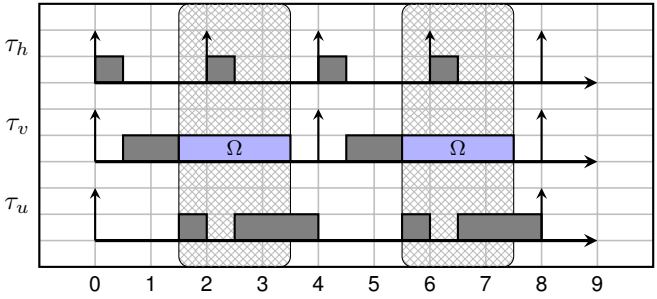

\begin{center}
\begin{RTGrid}[width=0.45\textwidth, height=3.5cm,nosymbols=1,nogrid=0,nonumbers=0,numbersize=\footnotesize]{3}{9}
\RTBox{1.5}{3.5}
\RTBox{5.5}{7.5}

\RowLabel{1}{$\tau_h$}
\TaskArrival{1}{0}
\TaskArrival{1}{2}
\TaskArrival{1}{4}
\TaskArrival{1}{6}
\TaskArrival{1}{8}

\TaskExecution{1}{0}{0.5}
\TaskExecution{1}{2}{2.5}
\TaskExecution{1}{4}{4.5}
\TaskExecution{1}{6}{6.5}

\RowLabel{2}{$\tau_v$}
\TaskArrival{2}{0}
\TaskArrival{2}{4}
\TaskArrival{2}{8}

\TaskExecution{2}{0.5}{1.5}
\TaskExecution[color=blue!30,execlabel=$\Omega$]{2}{1.5}{3.5}
\TaskExecution{2}{4.5}{5.5}
\TaskExecution[color=blue!30,execlabel=$\Omega$]{2}{5.5}{7.5}

\RowLabel{3}{$\tau_u$}
\TaskArrival{3}{0}
\TaskArrival{3}{8}
\TaskExecution{3}{1.5}{2}
\TaskExecution{3}{2.5}{4}
\TaskExecution{3}{5.5}{6}
\TaskExecution{3}{6.5}{8}

\end{RTGrid}
\end{center}
\caption{Attack effective window of $\tau_v$. Task parameters: $\tau_h = (0.5,2)$, and $\tau_u = (4,8)$, and  $\tau_v = (1,4)$ with attack effective window of length~$\Omega=2$. }
\label{fig:aew_example}
\end{figure}

An example of AEW is shown in Figure~\ref{fig:aew_example}. 
The window is associated with victim task $\tau_v$ and is marked in blue.
$\tau_h$ is a higher priority trusted task and $\tau_u$ is an untrusted task that might be an attacker. 
We define a window is covered when all its time slots are utilized by trusted tasks.
In this case a large part of $AEW$ is not covered and leaves place for untrusted task to execute. 
This is considered as unsafe.

The timing of the $AEW$ depends on the type of associated schedule-based attack. 
E.g., for anterior attack the $AEW$ will exist before the execution of the victim task, while for posterior attack the $AEW$ is after the execution of the victim task. 
In order to successfully carry out the attack, the attacker needs to execute during the $AEW$ following the execution of the victim task such that victim's secret can be stolen, corrupted, or overwritten. 

To summarize, the attacker considered in this paper is only able to penetrate the system through remote code execution on the target platform and gain device I/O access but can neither gain scheduling capabilities nor kernel privileges.
Hence, we assume the system kernel (including the scheduler) is secure from the manipulation of an attacker.
The attacker aims to successfully initiate a posterior schedule-based attack which means the attacker needs to execute during the $AEW$ for the chosen attack. 

In this paper we also consider a vendor oriented security model~\cite{pellizzoni2015generalized} where tasks from the same vendor are considered to be trusted task (as they share security designs) and as a result are less likely to be penetrated by an attacker.
All other tasks are considered untrusted and only untrusted tasks from other vendors have the potential to be the attacker in disguise.
Each task is assigned the minimum set of required capabilities following the principle of least privilege.
We assume a mixed criticality system where priorities of trusted and untrusted tasks can interleave.
There  is  only one victim task (denoted by~\(\tau_v\)) that carries  out security sensitive computation at the end of its execution, such as accessing important information in the cache or writing results to a buffer. 
There is no requirement on the relative priority between the $\tau_v$ and untrusted tasks. 

\section{Defense approaches}

\subsection{Philosophy}

The successful execution of an attacker task during $AEW$ is crucial to the success of the attack.
Hence, our defense focuses on using scheduling techniques to block all untrusted tasks from executing during $AEW$. 
To this end, we define two approaches: (a) paranoid approach and (b) trusted execution~approach. 

\subsection{\paranoid Approach}

A simple, brute-force approach would be to block all tasks from execution during $AEW$, say by using the system idle task to occupy this window. 
This would be equivalent to introducing the Flush task approach to prevent information leakage used by Mohan et al.~\cite{mohan2014real} and Pellizzoni et al.~\cite{pellizzoni2015generalized}.
This can fulfill our defense goal but at the cost of reducing the schedulability of the system. 
We consider this to be the base approach and is the conservative but safe approach. 

\subsection{Trusted execution approach}

Blocking all tasks from executing during the window wastes CPU cycles and reduces system utilization. 
The trusted execution approach that we propose would be blocking only untrusted tasks during $AEW$, since trusted tasks are considered~safe.

This method will improve the response time of all tasks compared to the base approach.
However, if trusted tasks cannot use up the entire window time or there are lower priority trusted tasks executing during the window, it can still block higher priority untrusted tasks and they may still miss their deadlines.
To solve this problem, we aim to answer the following questions:

\begin{itemize}
  \item How will response time of tasks change when using the trusted execution approach compared with the paranoid~approach?
  
  \item Given a set of trusted tasks, determine if all instances of the $AEW$ are covered?

  \item Is there a security-oriented scheduling policy that prioritizes window covering as much as possible while maintaining system schedulability?
\end{itemize}

\section{Analysis}
\label{analysis}

In this section, we first provide response time analysis for both paranoid and trusted execution scheduling approaches. 
Then we discuss in a unique condition how to determine if a given trusted task set can fully cover all instances of $AEW$ and what benefits it can bring. At last, we present a new scheduling policy that prioritizes the covering of $AEW$ while not affecting the system's schedulability. 

We assume in our analysis there is only one victim task~$\tau_v$ and its $AEW$ is not longer than its period,~\mbox{$\win < T_v$.}



\subsection{Response Time for Paranoid Approach}

We first consider the scheduling problem with the paranoid defense mechanism where none task is allowed to execute within $AEW$. 
The window can be modeled as a fixed non-preemptive region~\cite{Bril2007}.
To compute the safe bounds on the tasks' worst-case response times, we assume arbitrary phasing.
We first analyze the tasks with higher priorities than the victim task, then the tasks with lower priorities than the victim task, and finally, the victim task.

\textbf{Response time for higher priority task}
A task~$\tau_i\in hp(v)$  can be blocked by one attack effective window $B_i = \win$ in the worst case, and its worst-case response time is:
\begin{equation}
    R_i = C_i + \win + \sum_{j\in hp(i)} \left \lceil \dfrac{R_i}{T_j}  \right \rceil C_j
    \label{eq:rta_hp_paranoid}
\end{equation}


\textbf{Response time for lower priority task}
A task~$\tau_i \in lp(i)$  can be blocked by all instances of the victim task window:

\begin{equation}
    R_i = C_i + \sum_{j\in hp(i)} \left \lceil \dfrac{R_i}{T_j}  \right \rceil C_j + \left \lceil \dfrac{R_i}{T_v}  \right \rceil \win
    \label{eq:rta_lp_paranoid}
\end{equation}

\textbf{Response time for victim task}
The victim task analysis follows the principle of the non-preemptive response time analysis~\cite{Bril2007,Davis2007}: the non-preemptive window can overlap with the victim's task next instance or  block higher priority tasks deferring their execution into  the victim's task next instance.
The worst-case response time for $\tau_v$ will occur during $\tau_v$ busy period~$L_v$  (the longest time interval that the processor is occupied without idle time with tasks that have priorities higher than or equal to $\tau_v$~\cite{Lehoczky}) given by the least positive integer satisfying the  following relation:
\begin{equation}
    L_v =  \sum_{j\in hp(v)} \left \lceil \dfrac{L_v}{T_j}  \right \rceil C_j +   \left \lceil \dfrac{L_v}{T_v}  \right \rceil \left (C_v + \win \right )
\end{equation}
Let $r_{v,k}= (k-1) \cdot T_v$ be the $k$-th release time of the victim task~$\tau_v$ and $f_{v,k}$ its worst-case finish time ($finish\_time = release\_time + response\_time$):

\begin{equation}
    f_{v,k} =  \sum_{j\in hp(v)}\left\lceil \dfrac{f_{v,k}}{T_j} \right\rceil C_j +
    \tikzmarknode{b}{\highlight{NavyBlue}{
    $(k-1) \cdot \win$}} + k \cdot C_v
\end{equation}

\begin{figure}[t]
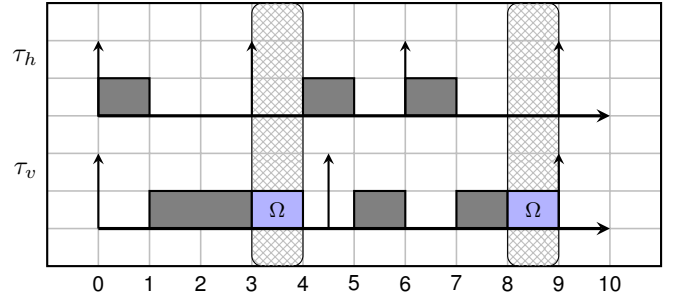

\begin{center}
\begin{RTGrid}[width=0.45\textwidth,height=3.5cm,nosymbols=1,nogrid=0,nonumbers=0,numbersize=\footnotesize]{2}{10}
\RTBox{3}{4}
\RTBox{8}{9}

\RowLabel{1}{$\tau_{h}$}
\TaskArrival{1}{0}
\TaskArrival{1}{3}
\TaskArrival{1}{6}
\TaskArrival{1}{9}

\TaskExecution{1}{0}{1}
\TaskExecution{1}{4}{5}
\TaskExecution{1}{6}{7}

\RowLabel{2}{$\tau_v$}
\TaskArrival{2}{0}
\TaskArrival{2}{4.5}
\TaskArrival{2}{9.0}

\TaskExecution{2}{1}{3}
\TaskExecution[color=blue!30,execlabel=$\win$]{2}{3}{4}

\TaskExecution{2}{5}{6}
\TaskExecution{2}{7}{8}
\TaskExecution[color=blue!30,execlabel=$\win$]{2}{8}{9}

\end{RTGrid}
\end{center}
\caption{Victim task response time analysis under Paranoid Approach: the victim task~$\tau_v=(2,4.5)$ with $\Omega=1$, and higher priority task $\tau_{h}=(1,3)$. Note that $R_{v,1}=3$ and $R_{v,2}=3.5$.}
\label{fig:paranoid_critical_instant}
\end{figure}

\begin{tikzpicture}[overlay,remember picture,>=stealth,nodes={align=left,inner ysep=1pt},<-]
    \path (b.north) ++ (0cm,2.5em) node[anchor=north west,color=NavyBlue!85] (mitext){\textsf{\footnotesize sum of  windows}};
    \draw [color=NavyBlue!85](b.north) |- ([xshift=-0.3ex,color=NavyBlue]mitext.south east);
\end{tikzpicture}
Its worst-case response time is calculated as the maximum of the response times of all instances.
\begin{equation}
    R_v = \max_k \left \{ f_{v,k} - r_{v,k} \right \}
\end{equation}
where $k: 0 < k \leq  L_v/T_v$.   
Figure~\ref{fig:paranoid_critical_instant} shows an example of the victim task response time~analysis.

\subsection{Response Time for Trusted Execution Approach}
The trusted execution approach allows the trusted tasks to execute within the victim's window.
We introduce the response time analysis for the trusted and untrusted tasks under the trusted execution approach. 
Our analysis assumes arbitrary phasing except for the victim task that, to simplify the presentation, has no initial offset~($\phi_v=0$).

\begin{figure}[t]
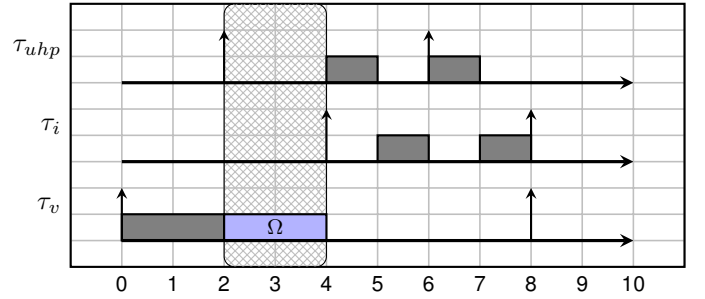

\begin{center}
\begin{RTGrid}[width=0.45\textwidth,height=3.5cm,nosymbols=1,nogrid=0,nonumbers=0,numbersize=\footnotesize]{3}{10}

\RTBox{2}{4}

\RowLabel{1}{$\tau_{uhp}$}
\TaskArrival{1}{2}
\TaskArrival{1}{6}

\TaskExecution{1}{4}{5}
\TaskExecution{1}{6}{7}

\RowLabel{2}{$\tau_{i}$}
\TaskArrival{2}{4}
\TaskArrival{2}{8}

\TaskExecution{2}{5.0}{6.0}
\TaskExecution{2}{7.0}{8.0}

\RowLabel{3}{$\tau_v$}
\TaskArrival{3}{0}
\TaskArrival{3}{8}

\TaskExecution{3}{0}{2}
\TaskExecution[color=blue!30,execlabel=$\win$]{3}{2}{4}

\end{RTGrid}
\end{center}
\caption{Critical instant for higher priority trusted tasks under Trusted Execution Approach: the victim task~$\tau_v=(2,8)$ with $\Omega=2$, trusted task under analysis $\tau_i=(2,4)$, and untrusted higher priority task~$\tau_{uhp}=(1,4)$.}
\label{fig:trusted_hp_rta}
\end{figure}

\textbf{Response time for higher priority trusted task and victim~task}
Task~$\tau_i \in thp(v)$ can experience interference from the higher priority (trusted and untrusted) tasks. 
The victim task can block the untrusted tasks, and consequently, the time between the start of the execution of the first and the second untrusted task instances can be less than its period.
Figure~\ref{fig:trusted_hp_rta} illustrates such a situation. 
Two instances of untrusted higher priority task $\tau_{uhp}$ interfere with the task under analysis~$\tau_i$ in the time interval from 4 to 8. 
To account for this, $\tau_{uhp}$ delayed by $\win$ from the previous instance should be considered.
The worst-case response time of trusted task~$\tau_i$ with a priority higher than~$\tau_v$ is the least positive integer that satisfies  the following recurrent equation:
\begin{equation}
    R_i = C_i + \sum_{j\in thp(i)} \left \lceil \dfrac{R_i}{T_j}  \right \rceil C_j + \sum_{j\in uhp(i)} \left \lceil \dfrac{R_i + \win}{T_j}  \right \rceil C_j
    \label{eq:rta_trusted_hp}
\end{equation}
We can use the above formula to safely upper bound the victim task's worst-case response time ~($i=v$).

\textbf{Response time for higher priority untrusted task}
Task~$\tau_i \in uhp(v)$, besides the interference from the higher priority (trusted  and  untrusted) tasks, can be blocked at most once by the victim's window.
Since the trusted tasks can execute during the victim's window, the window time will not contribute to additional blocking. The critical instance for $\tau_i \in uhp(v)$ happens when: i) $\tau_i$ and other $uhp(i)$ are released at the beginning of the window, ii) trusted higher priority tasks $thp(i)$ interfering with the execution of~$\tau_i$ are all released right after the end of the window. 
\begin{equation}
    R_i = C_i +  \win + \sum_{j\in thp(i)} \left \lceil \dfrac{R_i-\win}{T_j}  \right \rceil C_j +  \sum_{j\in uhp(i)} \left \lceil \dfrac{R_i}{T_j}  \right \rceil C_j
    \label{eq:rta_hp_untrusted}
\end{equation}


\textbf{Response time for lower priority untrusted task}
Task~$\tau_i \in ulp(v)$ is subject to interference from the higher priority tasks (trusted and untrusted) and from the window that is non-preemptive for untrusted tasks. The window is activated every time the victim task completes its execution.

\begin{figure}[t]
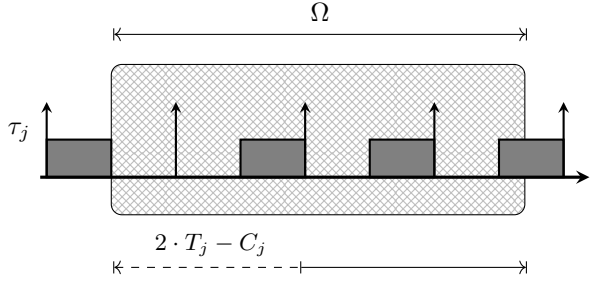

\begin{center}
\begin{RTGrid}[width=0.45\textwidth,height=2.0cm,nogrid=1, nosymbols=1, nonumbers=1]{1}{17}
    \RTBox{2.2}{15}
    \TaskArrival{1}{0.2}
    \TaskArrival{1}{4.2}
    \TaskArrival{1}{8.2}
    \TaskArrival{1}{12.2}
    \TaskArrival{1}{16.2}
    \TaskExecution{1}{0.2}{2.2}
    \TaskExecution{1}{6.2}{8.2}
    \TaskExecution{1}{10.2}{12.2}
    \TaskExecution{1}{14.2}{16.2}

    \draw node at (0.15,1.1) {$\tau_j$};
    
    \draw node at (4.15,2.7) {$\win$}; 
    \draw [|<->|] (1.4, 2.4) -- (6.9, 2.4);

    \draw node at (2.7,-0.4) {\small $2 \cdot T_j - C_j$};
    \draw [dashed, |<-|] (1.4, -0.7) -- (3.9, -0.7); 
    \draw [->|] (3.9, -0.7) -- (6.9, -0.7);

\end{RTGrid}
\end{center}
\caption{Trusted task~$\tau_j$ minimal amount of execution within window~$\win$.}
\label{fig:min_execution_in_window}
\end{figure}

The jobs of trusted higher priority task $\tau_{j} \in thp(i)$  executed within the window can be excluded from the set of the interfering jobs.
We derive a lower bound on the minimal amount of~$\tau_j$ execution within the window of length~$\win$.
Figure~\ref{fig:min_execution_in_window} illustrates our approach.
We assume that every instance of~$\tau_j$  executes for its worst-case execution time. The first~$\tau_j$ instance starts at its release, and every subsequent instance starts at~$C_j$ before its deadline.
The window starts right after the end of the first~$\tau_j$ instance.
Such conditions minimize the total execution of~$\tau_j$ within the window (shifting the window to the left or the right cannot decrease the total workload executed within).
\begin{equation}
  W_{\min}(j) \; = \; \max \left (0, \left \lceil \dfrac{ \win - 2 \cdot T_j + C_j }{T_j} \right \rceil \right )  C_j \end{equation}

The proposed approach is similar to the response time analysis for the polling servers~\cite{ButtazzoBook}.
We acknowledge that the bound is not tight in general.

The worst-case response time of task~$\tau_i \in ulp(v)$ is the least positive integer of the following recurrence:

\vspace{1.2\baselineskip}
\begin{equation}
    R_i = C_i + \tikzmarknode{a}{\highlight{OliveGreen}{$\sum_{j\in hp(i)} \left \lceil \dfrac{R_i}{T_j} \right \rceil C_j$}} + 
    \tikzmarknode{b}{\highlight{NavyBlue}{$\left \lceil \dfrac{R_i}{T_v}  \right \rceil \mathcal{U}_i$}}
    \label{eq:rta_lp_untrusted}
\end{equation}
\begin{tikzpicture}[overlay,remember picture,>=stealth,nodes={align=left,inner ysep=1pt},<-]
    \path (a.north) ++ (0,1.2em) node[anchor=south east,color=OliveGreen!85] (scalep){\textsf{\footnotesize preempt by hp(v)}};
    \draw [color=OliveGreen!85](a.north) |- ([xshift=-0.3ex,color=OliveGreen]scalep.south west);
    \path (b.north) ++ (-0.5cm,2.2em) node[anchor=north west,color=NavyBlue!85] (mitext){\textsf{\footnotesize blocked by AEW}};
    \draw [color=NavyBlue!85](b.north) |- ([xshift=-0.3ex,color=NavyBlue]mitext.south east);
\end{tikzpicture}
where the upper bound on the  uncovered part of the window~is: 
\begin{equation}
    \mathcal{U}_i = \max \left (0,  \win - \sum_{j\in thp(i)} W_{\min}(j) \right )
\end{equation}

\textbf{Response time for lower priority trusted task}
We consider now task~$\tau_i \in tlp(v)$.
This task can benefit from the remaining window time in $AEW$ during its execution. We first evaluate the minimal amount of accumulated window time over time interval~$t$. 
Then we evaluate how much of this time might be taken by the other higher priority trusted tasks.
The remaining part of the trusted time can be used~by~$\tau_i$.

We calculate a lower bound on the minimal amount of trusted execution $\alpha(t)$ over a generic time interval of length~\mbox{$t>0$}.
Depending on the attack effective window duration, 
we can distinguish two cases:
i)~\mbox{$\win < T_v - R_v$}, and
ii)~$\win \in (T_v-R_v,T_v)$.

\begin{figure}[htb]
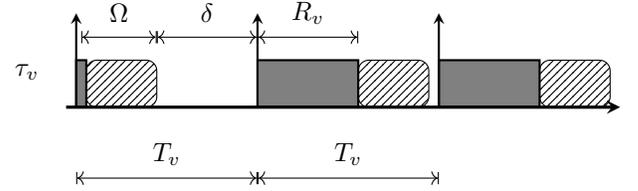

\begin{center}
\begin{RTGrid}[width=0.48\textwidth,height=2.5cm,nogrid=1, nosymbols=1, nonumbers=1]{1}{11}
    \TaskArrival{1}{0.2}
    \TaskArrival{1}{3.8}
    \TaskArrival{1}{7.4}
    
    \TaskParam{1}{0.2}{3.8}{$T_v$}
    
    \TaskParam{1}{3.8}{7.4}{$T_v$}

    \TaskExecution{1}{0.2}{0.4}
    \TaskExecution{1}{7.4}{9.4}

    \TaskExecution{1}{3.8}{5.8}
    \TaskParam{0}{3.8}{5.8}{$R_v$}
    
    
    

    \draw node at (0.15,1.1) {$\tau_v$};

   
    

    \TaskExecutionH{1}{0.4}{1.8}
    \TaskParam{0}{0.3}{1.8}{$\win$}
    \TaskParam{0}{1.8}{3.8}{$\delta$}
    
    \TaskExecutionH{1}{5.8}{7.2}
    \TaskExecutionH{1}{9.4}{10.8}
\end{RTGrid}
\end{center}
\caption{Non-overlapping  trusted execution  for~\mbox{$\win < T_v - R_v$}. }
\label{fig:min_trusted_execution_ii}
\end{figure}

In the first case, \mbox{$\win < T_v - R_v$}, the windows from two consecutive victims' jobs cannot overlap. Thus, within each victim task period, there is $\win$ trusted execution time.
\begin{equation}
    \alpha(t) \; = \; \max \left (0,  \left \lfloor \frac{t - \delta }{T_v} \right \rfloor \right ) \win 
\end{equation}
Variable $\delta$ is the maximal time from the end of the attack effective window to the next victim task release (this happens when the victim tasks finish immediately at its release time).
\begin{equation}
\delta =  T_v - \win 
\end{equation}
Figure~\ref{fig:min_trusted_execution_ii} shows an example of the victim task with non-overlapping windows and illustrates the above parameters.

\begin{figure}[htb]
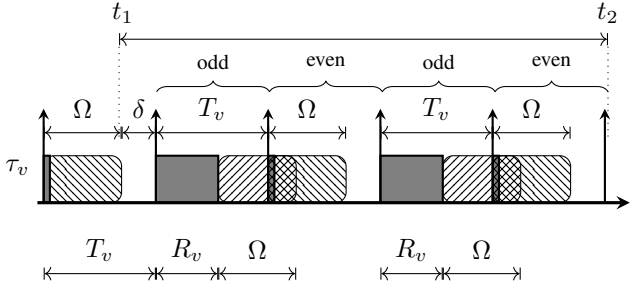

\begin{center}
\begin{RTGrid}[width=0.48\textwidth,height=2.5cm,nogrid=1, nosymbols=1, nonumbers=1]{1}{19}
    \TaskArrival{1}{0.2}
    \TaskArrival{1}{3.8}
    \TaskArrival{1}{7.4}
    \TaskArrival{1}{11.0}
    \TaskArrival{1}{14.6}
    \TaskArrival{1}{18.2}
    \TaskExecutionL{1}{0.2}{2.7}
    \TaskParam{0}{0.2}{2.7}{$\win$}
    
    \TaskParam{0}{2.7}{3.8}{$\delta$}
    \TaskParam{1}{0.2}{3.8}{$T_v$}

    \TaskExecution{1}{0.2}{0.4}
    \TaskExecution{1}{7.4}{7.6}
    \TaskExecution{1}{14.6}{14.8}
    
    \TaskExecution{1}{3.8}{5.8}
    \TaskParam{1}{3.8}{5.8}{$R_v$}
    \TaskExecutionH{1}{5.8}{8.3}
    \TaskParam{1}{5.8}{8.3}{$\win$}
    \TaskParam{0}{3.8}{7.4}{$T_v$}
    \TaskExecutionL{1}{7.4}{9.9}
    \TaskParam{0}{7.4}{9.9}{$\win$}
    
    \draw [decorate,decoration={brace,amplitude=4pt},xshift=0pt,yshift=0pt]
(2.0,2.1) -- (3.5,2.1) node [black,midway,xshift=-0.0cm, yshift=0.4cm] 
{\footnotesize odd};

    \draw [decorate,decoration={brace,amplitude=4pt},xshift=0pt,yshift=0pt]
(3.5,2.1) -- (5.0,2.1) node [black,midway,xshift=-0.0cm, yshift=0.4cm] 
{\footnotesize even};

    \draw [decorate,decoration={brace,amplitude=4pt},xshift=0pt,yshift=0pt]
(5.0,2.1) -- (6.5,2.1) node [black,midway,xshift=-0.0cm, yshift=0.4cm] 
{\footnotesize odd};

    \draw [decorate,decoration={brace,amplitude=4pt},xshift=0pt,yshift=0pt]
(6.5,2.1) -- (8.0,2.1) node [black,midway,xshift=-0.0cm, yshift=0.4cm] 
{\footnotesize even};
    
    \TaskExecution{1}{11}{13}
    \TaskParam{1}{11}{13}{$R_v$}
    \TaskParam{0}{11.0}{14.6}{$T_v$}
    \TaskExecutionH{1}{13.0}{15.5}
    \TaskParam{1}{13}{15.5}{$\win$}
    \TaskExecutionL{1}{14.6}{17.1}
    \TaskParam{0}{14.6}{17.1}{$\win$}

    \draw node at (0.15,1.1) {$\tau_v$};

   
   \draw node at (1.55,3.15) {$t_1$};
   \draw node at (8,3.15) {$t_2$}; 
   
   \draw [|<->|] (1.5, 2.8) -- (8.0, 2.8);
   \draw [dotted] (1.5,2.7) -- (1.5,1.5);
   \draw [dotted] (8.0,2.7) -- (8.0,1.5);
    
   
\end{RTGrid}
\end{center}
\caption{Minimal amount of trusted execution time for $\win > T_v -R_v$. }
\label{fig:min_trusted_execution_iii}
\end{figure}

In the second case, $\win \in (T_v-R_v,T_v)$, the windows from two consecutive victim jobs can overlap. Such overlapping leads to less trusted execution time.
Figure~\ref{fig:min_trusted_execution_iii} illustrates this case.
The reproduced schedule leads to the minimal time budget reserved for the execution of the trusted tasks within the time interval $[t_1,t_2]$ where $t_2 > t_1$ are time instants.
The time instant $t_1$ coincides with the end of the AEW that follows the victim  job executed instantaneously at its release time.
The amount of the trusted execution can be then minimal. To minimize the amount of the trusted execution, every two instances of the victim task~$\tau_v$ have overlapping windows. 
The first victim task instance within the interval $[t_1,t_2]$ terminates at its worst-case finishing time while the second one at its release. 
We will designate the first job of such a pair as the \emph{odd} job and the second one as the \emph{even} job.
The total trusted  execution time for each pair is~\mbox{$T_v - R_v + \win$.}

The \emph{odd} $\tau_v$ job gives rise to $T_v-R_v$ trusted execution time within its period. We cover in $[t_1,t_2]$ the period of the first \emph{odd} job after $T_v + \delta$, and then every $2\cdot T_v$, the next \emph{odd} job's period is covered.

\vspace{-0.4\baselineskip}

\begin{equation}
    \alpha_{odd}(t) \; = \; \max \left (0,
    \tikzmarknode{a}{\highlight{OliveGreen}{$\left \lfloor \dfrac{t + T_v-\delta}{2T_v} \right \rfloor$}} 
    \right ) 
    \tikzmarknode{b}{\highlight{NavyBlue}
    {$\left (T_v - R_v  \right )$}}
\end{equation}
\begin{tikzpicture}[overlay,remember picture,>=stealth,nodes={align=left,inner ysep=1pt},<-]
    \path (a.north) ++ (0.6cm,1.2em) node[anchor=south east,color=OliveGreen!85] (scalep){\textsf{\footnotesize \# of odd instances}};
    \draw [color=OliveGreen!85](a.north) |- ([xshift=-0.3ex,color=OliveGreen]scalep.south west);
    \path (b.north) ++ (-0.9cm,2.5em) node[anchor=north west,color=NavyBlue!85] (mitext){\textsf{\footnotesize trusted time budget}};
    \draw [color=NavyBlue!85](b.north) |- ([xshift=-0.3ex,color=NavyBlue]mitext.south east);
\end{tikzpicture}


The \emph{even} $\tau_v$ job gives rise to $\win$ trusted execution time within its period. We cover in $[t_1,t_2]$ the period of the first \emph{even} job after $\delta + 2 \cdot T_v$, and then every $2 \cdot T_v$, the next \emph{even} job's period is covered. 
\begin{equation}
    \alpha_{even}(t) \; = \; \max \left (0, \left \lfloor \dfrac{t-\delta}{2T_v} \right \rfloor  \right )  \win
\end{equation}

Putting it all together, the minimal amount of trusted execution $\alpha(t)$ can be lower bounded for $\win \in (T_v-R_v,T_v)$~as:
\begin{equation}
    \alpha(t) \; = \; \alpha_{even}(t) \; + \; \alpha_{odd}(t) 
\end{equation}


The derived bounds have simple expressions, thereby simplifying the analysis. 
However, the bounds are not tight. In particular, we account only for the trusted execution within the victim task periods that entirely fit the time interval.

We compute the amount of processing time reserved for a trusted task in any time interval.
During the time reserved for the trusted execution, task~$\tau_i$ contends for the processor only with trusted higher priority tasks.
We estimate the maximal amount of trusted execution time that might be reclaimed~by~$thp(i)$.

\begin{figure}[htb]
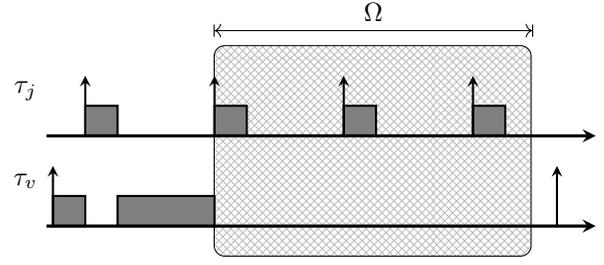

\begin{center}
\begin{RTGrid}[width=0.45\textwidth,height=2.8cm,nogrid=1, nosymbols=1, nonumbers=1]{2}{17}
    \RTBox{5.19}{15}
    \TaskArrival{1}{1.2}
    \TaskArrival{1}{5.2}
    \TaskArrival{1}{9.2}
    \TaskArrival{1}{13.2}

    \TaskExecution{1}{1.2}{2.2}
    \TaskExecution{1}{5.2}{6.2}
    \TaskExecution{1}{9.2}{10.2}
    \TaskExecution{1}{13.2}{14.2}
    
    \TaskArrival{2}{0.2}
    \TaskArrival{2}{15.8}
    \TaskExecution{2}{0.2}{1.2}
    \TaskExecution{2}{2.2}{5.2}

    \draw node at (0.15,2.2) {$\tau_j$};
    
    \draw node at (0.15,1.0) {$\tau_v$};
    
    \draw node at (4.775, 3.25) {$\win$}; 
    \draw [|<->|] (2.65, 3.0) -- (6.9, 3.0);


\end{RTGrid}
\end{center}
\caption{Higer priority trusted task~$\tau_j\in thp(v)$ maximal amount of execution within a~single attack effective window~$\win$ of task~$\tau_v$.}
\label{fig:max_trusted_execution}
\end{figure}

Since we assume the \emph{Rate Monotonic} priority assignment, trusted tasks with higher priorities than the victim task~$\tau_v$ have shorter periods than the victim and might be therefore released multiple times during~$AEW$. 
However, the first task $\tau_j\in thp(v)$ instance within the window must be released after the window's beginning.
Otherwise, $\tau_j$ could preempt $\tau_v$ (or some other $hp(v)$ task)  and execute before task~$\tau_v$ ends. Figure~\ref{fig:max_trusted_execution} shows task $\tau_j\in thp(v)$ executing within the attack effective window.
The maximal amount of trusted execution time reclaimed by~$\tau_j\in thp(v)$ within a~single window~$\win$ can be upper bounded by:
\begin{equation}
    W_{\max}(j) = \min \left( \win, \left \lceil  \dfrac{\win}{T_j} \right \rceil  C_j \right) 
\end{equation}
If $\alpha(t)$ is the minimal amount of the trusted execution over a generic time interval of length $t>0$, then a higher priority trusted task~$\tau_j \in thp(v)$ cannot use more trusted execution budget~than:
\begin{equation}
 \beta_j(t) =  \left \lceil \dfrac{\alpha(t)}{\win} \right \rceil  W_{max}(j) 
\end{equation}

\begin{figure}[h]
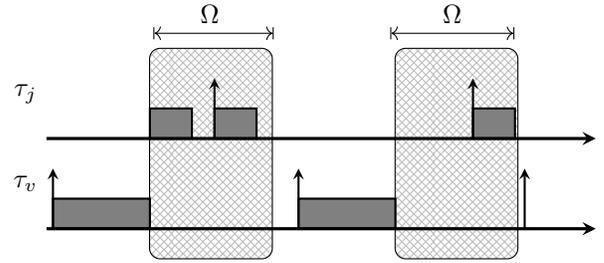

\begin{center}
\begin{RTGrid}[width=0.45\textwidth,height=2.8cm,nogrid=1, nosymbols=1, nonumbers=1]{2}{17}
    \RTBox{3.19}{6.99}
    \RTBox{10.79}{14.59}
    \TaskArrival{1}{5.2}
    \TaskArrival{1}{13.2}

    \TaskExecution{1}{13.2}{14.5}
    \TaskExecution{1}{3.2}{4.5}
    \TaskExecution{1}{5.2}{6.5}
    
    \TaskArrival{2}{0.2}
    \TaskArrival{2}{7.8}
    \TaskArrival{2}{14.8}
    \TaskExecution{2}{0.2}{3.2}
    
    \TaskExecution{2}{7.8}{10.8}

    \draw node at (0.15,2.2) {$\tau_j$};
    
    \draw node at (0.15,1.0) {$\tau_v$};
    
    \draw node at (2.6, 3.25) {$\win$}; 
    \draw [|<->|] (1.85, 3.0) -- (3.45, 3.0);
    
    \draw node at (5.825, 3.25) {$\win$}; 
    \draw [|<->|] (5.0, 3.0) -- (6.65, 3.0);


\end{RTGrid}
\end{center}
\caption{Lower priority trusted task~$\tau_j\in tlp(v)$ maximal amount of execution during a sequence of task~$\tau_v$ attack effective windows~$\win$.}
\label{fig:max_trusted_execution_lp}
\end{figure}

On the other hand, trusted tasks with lower priorities than the victim  task~$\tau_v$ (but with a priority higher than the task under analysis~$\tau_i$) have longer than or equal periods to the victim. 
Each job of such a lower priority task~$\tau_j\in tlp(v) \cap thp(i)$ can overlap with $AEW$. Two jobs can fit into the same $AEW$ if the first one finishes as late as possible and the next one as early as possible. Figure~\ref{fig:max_trusted_execution_lp} shows such a schedule. Within time interval of length $t>0$,  trusted task $\tau_j\in tlp(v) \cap thp(i)$   with  lower  priority  than  the  victim cannot reclaim more trusted execution time than:
\begin{equation}
 \beta_{j}(t) =  \left \lceil \dfrac{t + T_j - C_j}{T_j} \right \rceil \min (\win,C_j) 
\end{equation}

\begin{figure}[h]
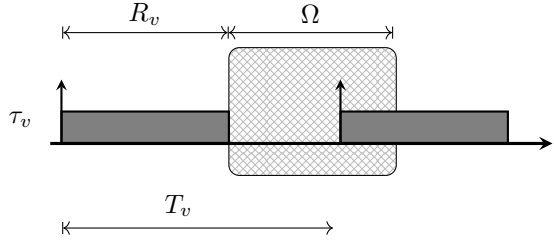

\begin{center}
\begin{RTGrid}[width=0.45\textwidth,height=1.7cm,nosymbols=1,nogrid=1,nonumbers=1,numbersize=\footnotesize]{1}{9}

\RTBox{3.2}{6.2}

\draw node at (4.2, 2.15) {$\win$}; 
\draw [|<->|] (3.1, 1.9) -- (5.3, 1.9); 

\draw node at (2.0, 2.15) {$R_v$};
\draw [|<->] (0.9, 1.9) -- (3.1, 1.9);

\draw node at (2.45, -0.4) {$T_v$};
\draw [|<->] (0.9, -0.7) -- (4.5, -0.7);

\draw node at (0.35,0.75) {$\tau_v$};
\TaskArrival{1}{0.2}
\TaskArrival{1}{5.2}

\TaskExecution{1}{0.2}{3.2}
\TaskExecution{1}{5.2}{8.2}


\end{RTGrid}
\end{center}
\caption{Victim task~$\tau_v$ running in attack effective window~$AEW$.}
\label{fig:chase_tail}
\end{figure}

Last but not least, a victim task~$\tau_v$ instance can execute during its previous instance of the window as shown in Figure~\ref{fig:chase_tail} and the time available for other lower priority trusted tasks could be further reduced. 
The amount of the victim task~$\tau_v$ execution that can overlap with its previous window can be upper bounded by:
\begin{equation*}
W_{\max}(v) = \max \left ( 0, \min \left (C_v,  R_v + \win - T_v \right ) \right )
\end{equation*}
Thus, the maximum amount of task~$\tau_v$ execution that can overlap with all windows in a generic time interval of length~$t>0$~is:
\begin{equation}
   \beta_v(t) = \left \lceil \dfrac{t}{T_v}  \right \rceil   W_{\max}(v)
\end{equation}

Putting it all together,  the amount of trusted processing time available for task~$\tau_i \in tlp(v)$ during a time interval~$t>0$ is lower bounded by:
\begin{equation}
   \lambda_i(t)  = \max \left ( 0, \alpha(t) -  \sum_{j\in thp(i) } \beta_j(t)  \right )
    \label{eq:available_trusted}
\end{equation}
The worst-case response time of a~trusted lower priority task~$\tau_i \in tlp(v)$ can be upper bounded by:

\vspace{0.5\baselineskip}
\begin{equation}
    R_i =  C_i + \sum_{j\in thp(i)} \left \lceil \dfrac{R_i}{T_j}  \right \rceil C_j + 
    \sum_{j\in uhp(i)}  \left \lceil \dfrac{R_i+\win}{T_j}  \right \rceil C_j
    - \tikzmarknode{b}{\highlight{NavyBlue}{$\lambda_i(t)$}} 
    \label{eq:rta_lp_trusted}
\end{equation}
\begin{tikzpicture}[overlay,remember picture,>=stealth,nodes={align=left,inner ysep=1pt},<-]
    \path (b.north) ++ (-2.5cm,2.5em) node[anchor=north west,color=NavyBlue!85] (mitext){\textsf{\footnotesize exec time in AEW}};
    \draw [color=NavyBlue!85](b.north) |- ([xshift=-0.3ex,color=NavyBlue]mitext.south west);
\end{tikzpicture}
or by the least positive integer value that satisfies:
\begin{equation}
    C_i - \lambda_i(R_i) \; \leq \; 0
    \label{eq:rta_lp_trusted_ii}
\end{equation}
Formula~(\ref{eq:rta_lp_trusted_ii}) can be satisfied when there is a sufficient amount of trusted execution time to fully execute task $\tau_i$.

\subsection{Window covering condition}
In previous subsections, we derived response time analysis under the proposed trusted execution approach. 
If all trusted tasks $\in hp(v)$ and can cover all instances of $AEW$, then the response time analysis for untrusted task will be much simpler.
In this subsection, we derive the conditions that a set of trusted tasks can cover all instances of $AEW$ in a special scenario.

We consider a \emph{harmonic} taskset (\emph{i.e.,} periods that pairwise divide each other) with constant execution times.
We also assume that tasks have criticality monotonic priorities (\emph{i.e.,}~all trusted  tasks  are  assigned a higher priority than the same).
Besides, we introduce $R_{i}^{+}$ as the worst-case response time of task $\tau_i$ when its worst-case execution time is inflated to $C_i+\epsilon$ where $\epsilon > 0$ is an infinitesimally small positive number or one clock cycle if the discrete-time model is used.

To check if there is any idle time for a length of $\win$ after the execution of victim task~$\tau_v$, we first provide a sufficient and necessary condition for only $\tau_i \in thp(v)$ to cover all instances of the window.
Task $\tau_v$ associated attack effective window $\win$ is always covered by $\tau_i \in thp(v)$ if and only if:
\begin{equation}
\label{eq:hp_task_cover}
    R_v^{+} \; \geq \; R_v + \win
\end{equation}


\begin{figure}[hbt]
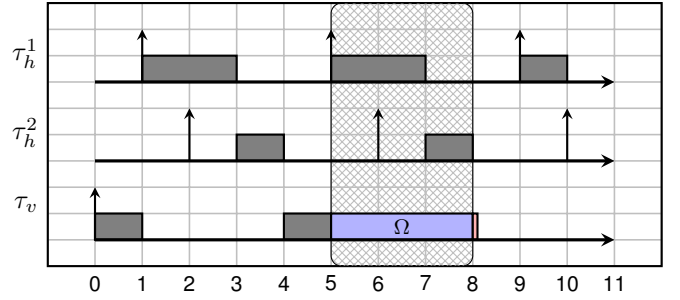

\begin{center}
\begin{RTGrid}[width=0.45\textwidth,height=3.5cm,nosymbols=1,nogrid=0,nonumbers=0,numbersize=\footnotesize]{3}{11}
\RTBox{5}{8}

\RowLabel{1}{$\tau_h^1$}
\TaskArrival{1}{1}
\TaskArrival{1}{5}
\TaskArrival{1}{9}
\TaskExecution{1}{1}{3}
\TaskExecution{1}{5}{7}
\TaskExecution{1}{9}{10}

\RowLabel{2}{$\tau_h^2$}
\TaskArrival{2}{2}
\TaskArrival{2}{6}
\TaskArrival{2}{10}
\TaskExecution{2}{3}{4}
\TaskExecution{2}{7}{8}

\RowLabel{3}{$\tau_v$}
\TaskArrival{3}{0}
\TaskExecution{3}{0}{1}
\TaskExecution{3}{4}{5}
\TaskExecution[color=blue!30,execlabel=$\win$]{3}{5}{8}
\TaskExecution[color=red!30]{3}{8}{8.1}

\end{RTGrid}
\end{center}
\caption{Higher priority tasks covering the attack effective window.}
    \label{fig:high_cover}
\end{figure}


As shown in Figure~\ref{fig:high_cover}, both $\tau_h^1$ and $\tau_h^2$ belong to a trusted task set and they both have higher priority than $\tau_v$. Task $\tau_v$ has the lowest priority among the three tasks and has an associated attack effective window $\win$ of 3 time units. 
With $C_v^{+}=C_v+\varepsilon$, the new response time of $\tau_v$ will be $8+\varepsilon$ ($\varepsilon$ is the red part in Figure~\ref{fig:high_cover}), which is larger than 8.
According to Formula~(\ref{eq:hp_task_cover}), the given task set can cover the window.

We consider how trusted tasks $\tau_i \in tlp(v)$ can fully cover the window.
Let $\tau_l \in tlp(v)$.
Task $\tau_v$ AEW is fully covered by $\tau_l$ if $R_l$ satisfies the following relation:
\begin{equation}
    R_l \; > \; I_{lv} + T_l - T_v + R_v + \win
\label{eq:lp_trusted_cover}
\end{equation}
where $I_{lv}$ is the difference in initial offset between $\tau_l$ and $\tau_v$. A positive $I_{lv}$ means $\tau_l$ starts earlier than $\tau_v$.
$R_l$ spans over all instances of $\tau_v$ within $\tau_l$ period.
$T_l - T_v + I_{lv}$ is the release time of the last $\tau_v$ instance within $\tau_l$ period.

Formula~(\ref{eq:lp_trusted_cover}) provides a sufficient condition for trusted task $\tau_l \in tlp(v)$ to cover the window.
However, it is also applicable to a trusted task sets: if there exist one task $\tau_i \in tlp(v)$ in a task set that satisfy Formula~(\ref{eq:lp_trusted_cover}), then this whole task set can fully cover the window.

 

Let  $\tau_l \in tlp(v)$. 
If the window is not already fully covered by $thp(v)$ 
then $\tau_v$ $AEW$ is fully covered by trusted tasks 
if and only if $R_l^{+}$ is larger than the last instance of the window in its period~$T_l$.
This provides a necessary and sufficient condition for lower priority to fully cover the window when higher priority tasks fail. 

To summarize the above findings:
Let $\tau_v$ be the victim task with associated attack effective window $\win$ and $\tau_l \in tlp(v)$  the trusted task with the lowest priority among all trusted tasks.
The window can be fully covered by the trusted tasks if and only if
\begin{equation}
    R_v^{+} \; \geq \; R_v +\win
\end{equation}
or 
\begin{equation}
\label{eq:fully_cover}
    R_l^{+} \; > \; I_{lv} + T_l - T_v + R_v + \win
\end{equation}

This provides a necessary and sufficient condition for a given trusted task set to cover the window entirely. 

In the following example, the response time of $\tau_l$ is 10 such that it cannot cover the last instance of the window. 
However, there is a high priority task~$\tau_h$ executing from 10-10.5, and $R_l^{+}$ will be larger than 10.5, leaving no idle time slot in the window.
According to Formula~(\ref{eq:fully_cover}), this task set can indeed fully cover the window.

\begin{figure}[hbt]
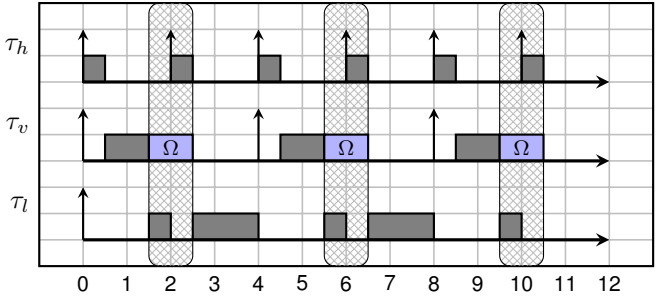

\begin{center}
\begin{RTGrid}[width=0.45\textwidth,height=3.5cm,nosymbols=1,nogrid=0,nonumbers=0,numbersize=\footnotesize]{3}{12}
\RTBox{1.5}{2.5}
\RTBox{5.5}{6.5}
\RTBox{9.5}{10.5}

\RowLabel{1}{$\tau_h$}
\TaskArrival{1}{0}
\TaskArrival{1}{2}
\TaskArrival{1}{4}
\TaskArrival{1}{6}
\TaskArrival{1}{8}
\TaskArrival{1}{10}

\TaskExecution{1}{0}{0.5}
\TaskExecution{1}{2}{2.5}
\TaskExecution{1}{4}{4.5}
\TaskExecution{1}{6}{6.5}
\TaskExecution{1}{8}{8.5}
\TaskExecution{1}{10}{10.5}

\RowLabel{2}{$\tau_v$}
\TaskArrival{2}{0}
\TaskArrival{2}{4}
\TaskArrival{2}{8}

\TaskExecution{2}{0.5}{1.5}
\TaskExecution[color=blue!30,execlabel=$\win$]{2}{1.5}{2.5}
\TaskExecution{2}{4.5}{5.5}
\TaskExecution[color=blue!30,execlabel=$\win$]{2}{5.5}{6.5}
\TaskExecution{2}{8.5}{9.5}
\TaskExecution[color=blue!30,execlabel=$\win$]{2}{9.5}{10.5}

\RowLabel{3}{$\tau_l$}
\TaskArrival{3}{0}
\TaskExecution{3}{1.5}{2}
\TaskExecution{3}{2.5}{4}
\TaskExecution{3}{5.5}{6}
\TaskExecution{3}{6.5}{8}
\TaskExecution{3}{9.5}{10}

\end{RTGrid}
\end{center}
\caption{Attack window covered with low and high priority tasks.}
    \label{fig:theorem_2}
\end{figure}




If a trusted task has the same priority as the victim task, the scheduler should break the tie by letting the victim task run first to provide better coverage. 

\subsection{Coverage oriented scheduling policy}
$AEW$ adds a new dimension to scheduling.
However, it is difficult to obtain information about $AEW$ and use it for scheduling because its length $\win$ is system dependant and attack dependant.
This section presents a new scheduling policy that is also unaware of this $AEW$ information but tries to provide best-effort security protection while maintaining system schedulability. 
The policy aims to maximize trusted task execution after the execution of victim task~$\tau_v$ for a period as long as possible.
The task to protect is marked as the victim task $\tau_v$.
For each task $\tau_i$, a maximum tolerable blocking time $B_i$ is calculated offline using exact RM schedulability analysis. 
Maximum tolerable blocking time, by definition, is the maximum time a task may be blocked from executing by lower priority tasks without eventually missing its deadlines.

For each scheduling instance, if there are tasks that have experienced a blocking time of its $B_i$, the scheduler will schedule the highest priority task from that group of tasks.
If the above condition is not valid, the scheduler will take different action based on whether $\tau_v$ is in the run queue and ready for execution. If $\tau_v$ is in the run queue, the scheduler will choose the highest priority task from $uhp(v)$.
If $uhp(v)$ is empty, the scheduler will choose $\tau_v$.
If $\tau_v$ is not in the run queue, the scheduler will choose the highest priority task from all trusted tasks until $\tau_v$ is inserted into the run queue again.
If $\tau_v$ is not in the run queue and there's no trusted task ready for execution, the scheduler will run the system idle task if no untrusted task has experienced a blocking time equal to $B_i$.
The pseudo-code for this scheduling policy is written in Algorithm~\ref{alg:policy}, and a comparison to RM is provided in Section~\ref{sec:sim_cover}.

This scheduling policy aims to push as many $uhp(v)$ as possible to execute before $\tau_v$ and pack as many trusted tasks as possible after the victim task.
Schedulability is guaranteed by scheduling tasks that cannot be blocked any further by other lower priority tasks. 
Although this scheduler cannot guarantee whether a window can always be fully covered or the length of the window can be covered, it tries to cover periods after $\tau_v$ as much as possible in a best effort way.
The scheduler achieves better security by sacrificing the response time of tasks since most of the tasks will finish close to their worst-case response~time.

\begin{algorithm} 
    \caption{Coverage oriented scheduling policy}
    \begin{algorithmic}[1]
    \State All task assigned priority $p_i$ according to RM
    \State $B_i$ is calculated for each task $\tau_i$ 
    \State For each scheduling instance
    \If{Some $\tau_i$ have been blocked for $B_i$}
         \State Select top task from these tasks
    \ElsIf{$\tau_v$ in run queue}
        \If{Some $\tau_i \in uhp(v)$ in run queue}
            \State Select top task in $uhp(v)$
        \Else
            \State Select $\tau_v$
        \EndIf
    \ElsIf{$\tau_v$ not in run queue}
        \If{Some trusted $\tau_i$ in run queue}
            \State Select top task from all trusted tasks
        \Else
            \State Select system idle task
        \EndIf
    \EndIf
    \end{algorithmic}
\label{alg:policy}
\end{algorithm}

\section{Implementation}
\label{sec:eval}
In this section, we describe how \emph{SchedGuard} was implemented in the Linux kernel before evaluating it in Section~\ref{sec:eval}.
To achieve the \emph{SchedGuard} functionality in the Linux kernel, we modified the kernel scheduler and made necessary changes to the cgroup interface as we choose to support containers.
Containers offer low-performance overhead, support for Linux-based OS, ease of porting software, and isolation enforced by namespace.
 They can be controlled through cgroups, making them compatible with vendor-oriented security models. 
The implementation of \emph{SchedGuard} assumes that all trusted tasks run in one container, while untrusted tasks run in one or several other containers.
This implementation targets a context of a uniprocessor system. 

  
  
  
  

\begin{figure}[tp]
    \centering
    \includegraphics[width=\linewidth]{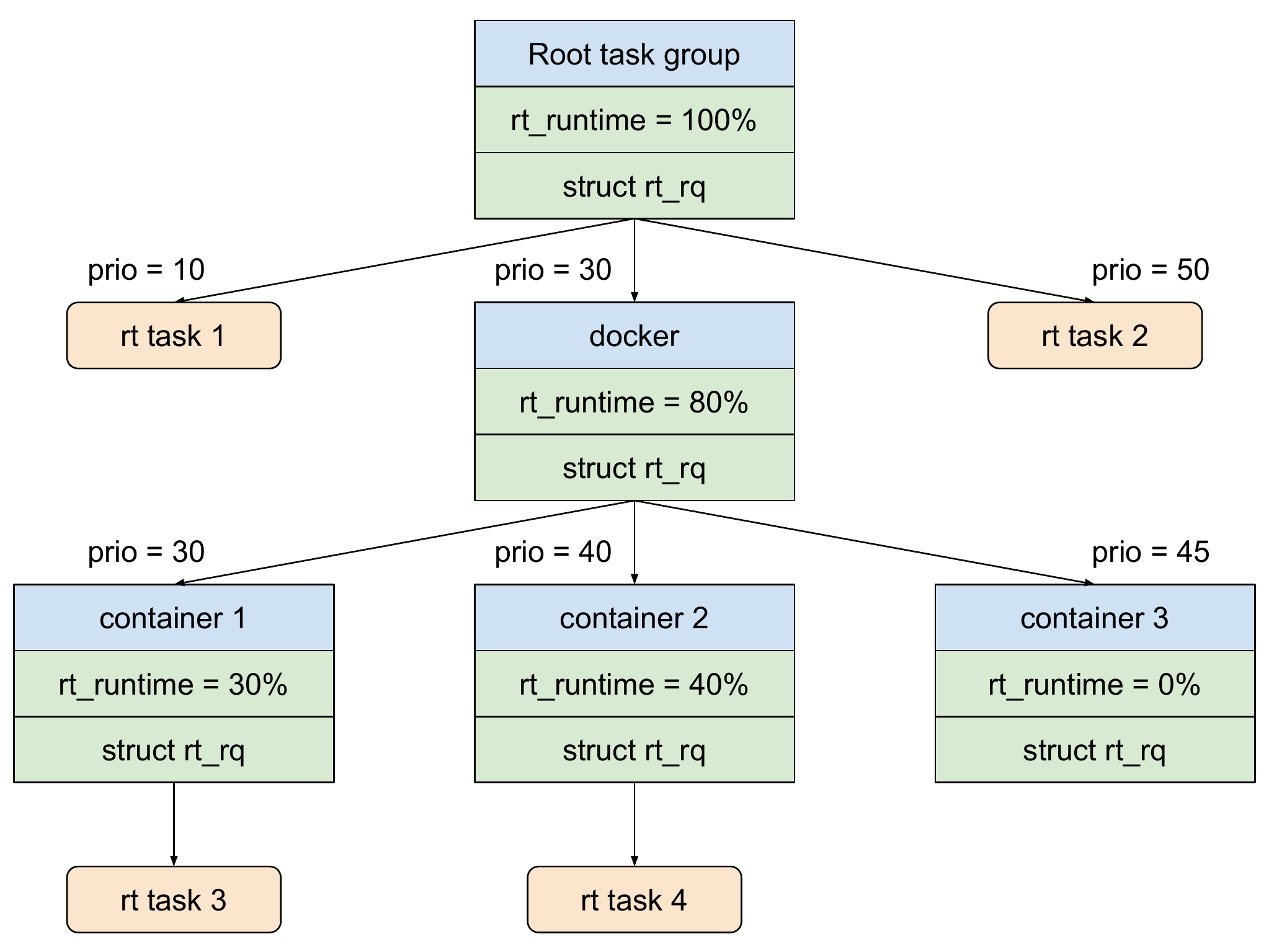}
    \caption{Docker cgroup structure.}
    \label{fig:docker}
\end{figure}

Linux cgroups are hierarchical groups that organize different resources for a collection of processes to perform resource allocation and monitoring. 
Examples include CPU, memory, device I/O, network, etc.
In this paper, we only discuss the components that are relevant to real-time scheduling on the CPU.
The CPU subsystem controls cgroup tasks access to the CPU.
It has a real-time bandwidth control feature that regulates the CPU real-time runtime (rt\_runtime) assigned for cgroup tasks. 
When the rt\_runtime of a cgroup is depleted, real-time tasks in this cgroup are stalled regardless of their priority until bandwidth replenishment in the next real-time period.
In the Linux kernel, there is a root task group that sits at the root of the cgroup hierarchy.
By default, all real-time bandwidth is assigned to this root task group, and any new cgroup can inherit rt\_runtime from its parent cgroup.
Take Docker, for example.
As shown in Figure~\ref{fig:docker}, Docker is a direct child of the root task group, and all containers are child cgroups of Docker.
As a result, the sum of rt\_runtime of all containers cannot be higher than Docker, which gets rt\_runtime from the root task group.
Each cgroup has a real-time run queue (rt\_rq) that stores information of real-time tasks in this cgroup. 
During real-time scheduling, the kernel always starts by searching in the root cgroup's rt\_rq for the highest priority real-time scheduling entity (sched\_rt\_entity), which can be either a task or cgroup.
If the selected sched\_rt\_entity is a cgroup, then the scheduler searches within the rt\_rq of this cgroup until it finds a task to execute.

To enable \emph{SchedGuard} blocking, one should first specify the protection window's length for the victim's cgroup.
In our extension of the cgroup implementation, this can be achieved using the cgroup file system by setting the cpu.window\_us attribute to a non-zero value.
The cpu.window\_us value is used to set the expiration time of the \emph{SchedGuard} hrtimer in the kernel.
To use the \emph{SchedGuard}, the victim task at the run time calls our newly added system call named cpu\_block right before calling yeild.
The cpu\_block ensures two functionalities: 1) it sets the kernel scheduler into protection mode; 2) it programs the \emph{SchedGuard} hrtimer to fire in the future.
In the protection mode, the kernel scheduler dequeues all rt\_rqs that have real-time tasks ready to execute except the rt\_rq of the victim's cgroup and the rt\_rq of root task group's as it may have real-time kernel tasks. 
Suppose there are no real-time tasks ready for execution from the victim's cgroup or the root task group during protection mode. In that case, the kernel scheduler will skip scheduling of all SCHED\_NORMAL tasks (normally handled by the CFS scheduler) and select the system idle task for running until the protection window is finished. 
When the \emph{SchedGuard} hrtimer expires, it reset the kernel scheduler back to normal mode and enqueues all dequeued~rt\_rqs.



  






\section{Experiments}
\label{experiments}

This section describes the experimental setup where we have demonstrated our proposed approach's results on a realistic platform, a radio-controlled rover (RC) car.
Moreover, we also provide the theoretical schedulability results of different defense approaches using synthetically generated workloads.



\subsection{Experimental Results on RC Car}

The computing unit on the RC car employs a Raspberry PI~4B.  
It has quad-core cortex A-72 cores capable of running at 1.5Ghz each and comes with Linux kernel 4.19 pre-installed. 
For our hardware experiments, we enabled only one core.
To validate our approach's effectiveness, we first show the results when \emph{SchedGuard} is used with a synthetic victim task, and then we show how it can protect the RC car's autopilot application. 
The simulation experiments run on a desktop environment and demonstrate the proposed security-oriented scheduling policy using a synthetically generated victim task.

\noindent\textbf{Defense against timing inference attack}

There are works such as \emph{ScheduLeak} \cite{chen2018scheduleak} that exploits scheduling side-channel information to reconstruct a periodic victim task initial offset (\emph{i.e.}, the arrival time) and case execution time.
With this information, an attacker can carry out an accurate timing-based attack without leaving any footprint.
\emph{SchedGuard} can affect the inference on execution time since it blocks the attacker task from obtaining any information during the protection window.

The defense is demonstrated in the following example.
The \emph{ScheduLeak} algorithm is used to infer the victim task initial offset $\tau_v$'s initial offset $a_v$ (i.e., the arrival time) and best case execution time $e_v$. 
The observer task from ScheduLeak is configured as a SCHED\_FIFO task with the lowest real-time priority in the system. 
The victim task is a periodic real-time task that runs with a 100ms period.
The measured average execution time and best case execution time for the victim task are 30ms and 19ms, respectively. 
We run only one periodic task (the victim task) in the system (excluding \emph{ScheduLeak} itself and kernel threads) as this increases the chance the inference can succeed.
The victim's period is passed to \emph{ScheduLeak} as it's a prerequisite condition for it to succeed.
To protect the victim task with \emph{SchedGuard}, the victim runs in a dedicated container alone, and a blocking window of 10ms is assigned. 
This container is assigned a rt\_runtime around 400ms over a period of 1000ms to make sure its execution is not affected by cgroup's RT throttling mechanism. 
The \emph{ScheduLeak} algorithm runs in a different container following the vendor-oriented security assumption, and the rest of the system's remaining  rt\_runtime (550ms) is assigned to it to increase its success rate.
After the victim starts execution, \emph{ScheduLeak} is invoked to run for 10 x victim's period following the original paper's recommendation.  



\begin{figure}[tp]
\centering
\includegraphics[width=\linewidth]{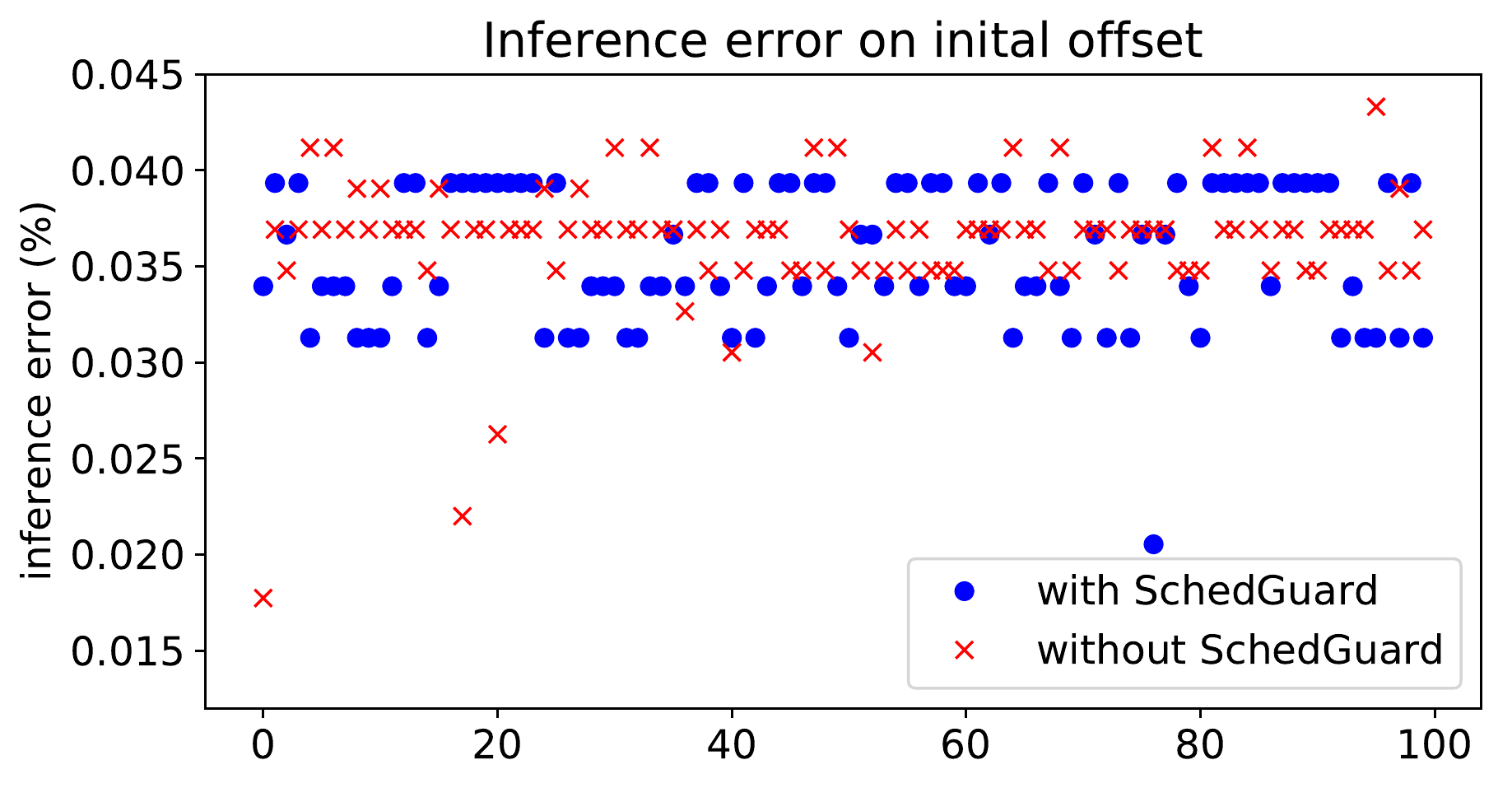}
\caption{ScheduLeak inference on task initial offset with and without SchedGuard.}
\label{fig:initial}
\end{figure}


\begin{figure}[tp]
\centering
\includegraphics[width=\linewidth]{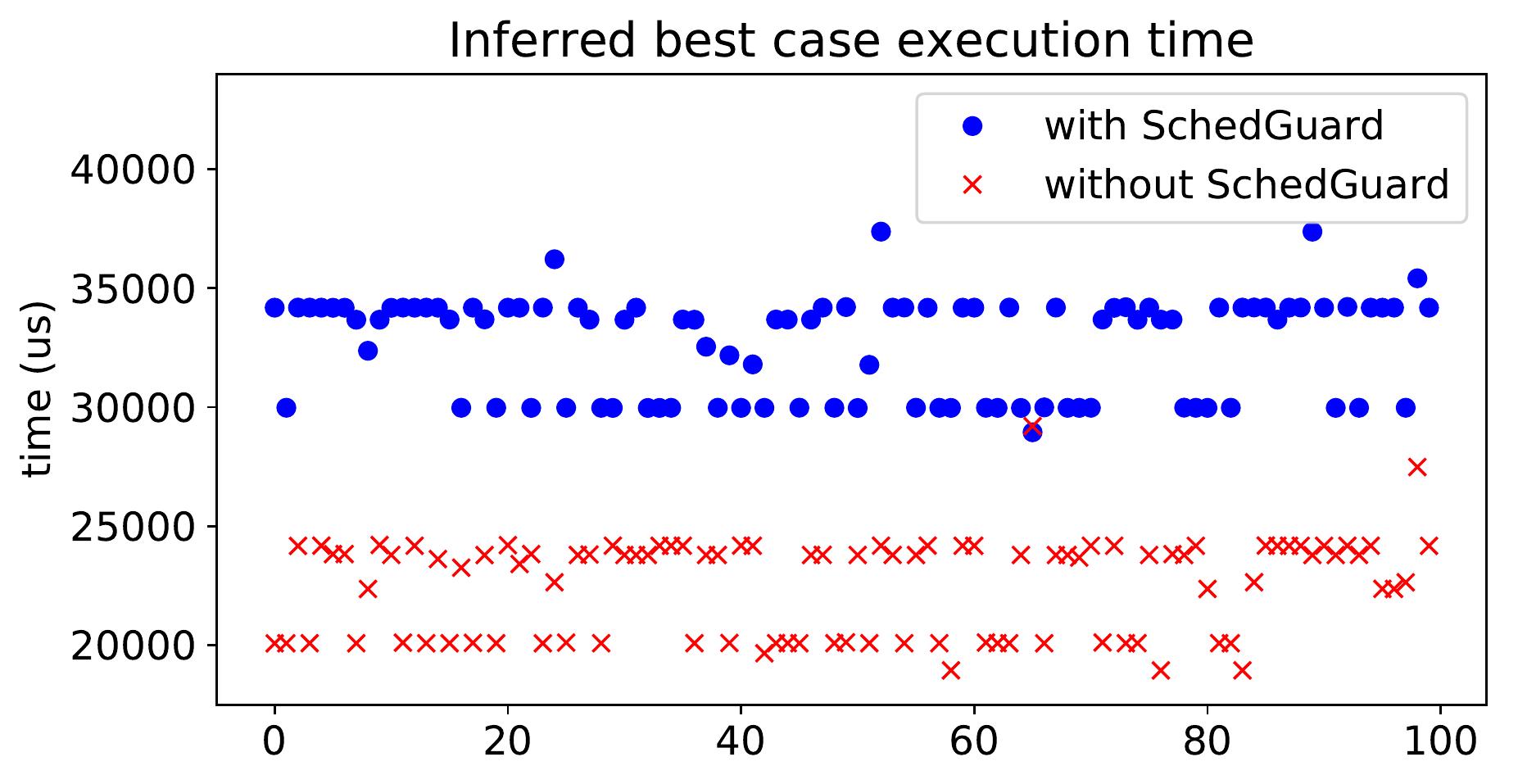}
\caption{ScheduLeak inference on victim task BCET with and without~SchedGuard.}
\label{fig:bcet}
\end{figure}

The \emph{ScheduLeak} algorithm is run 100 times for both \emph{SchedGuard} enabled and disabled cases. 
Inference results on the victim's initial offset and best case execution time are shown in Figure~\ref{fig:initial} and Figure~\ref{fig:bcet}.
Figure~\ref{fig:initial} shows the percentage error in victim task initial offset inference for both configurations.
ScheduLeak can derive a very accurate $a_v$ for the victim with only minor errors in both cases. 
This is because the \emph{SchedGuard} does not prevent the attacker from obtaining this~information. 

The inference results on the victim task's BCET are shown in Figure~\ref{fig:bcet} for both configurations.
The actual inference value instead of the percentage error is shown. 
The victim task has a true BCET of 19000us, while the majority of inference results fall between 20000us to 24000us when \emph{SchedGuard} is disabled.
When \emph{SchedGuard} is enabled with a 10ms ($\win=10ms$) protection window, most results range from 30000us to 34000us. 
Comparing with the no protection case, the difference is the protection window size.
This proves that \emph{SchedGuard} prevents the attacker from executing within 10ms after the victim finishes and gives the attacker an impression that the victim has a longer execution time. 
With this false execution time, the attacker will launch a posterior attack at the wrong moment.
If the protection window is longer than the attack effective window for that specific attack, the system is protected by \emph{SchedGuard}. 



\begin{figure}[tp]
    \centering
    \includegraphics[width=0.8\linewidth]{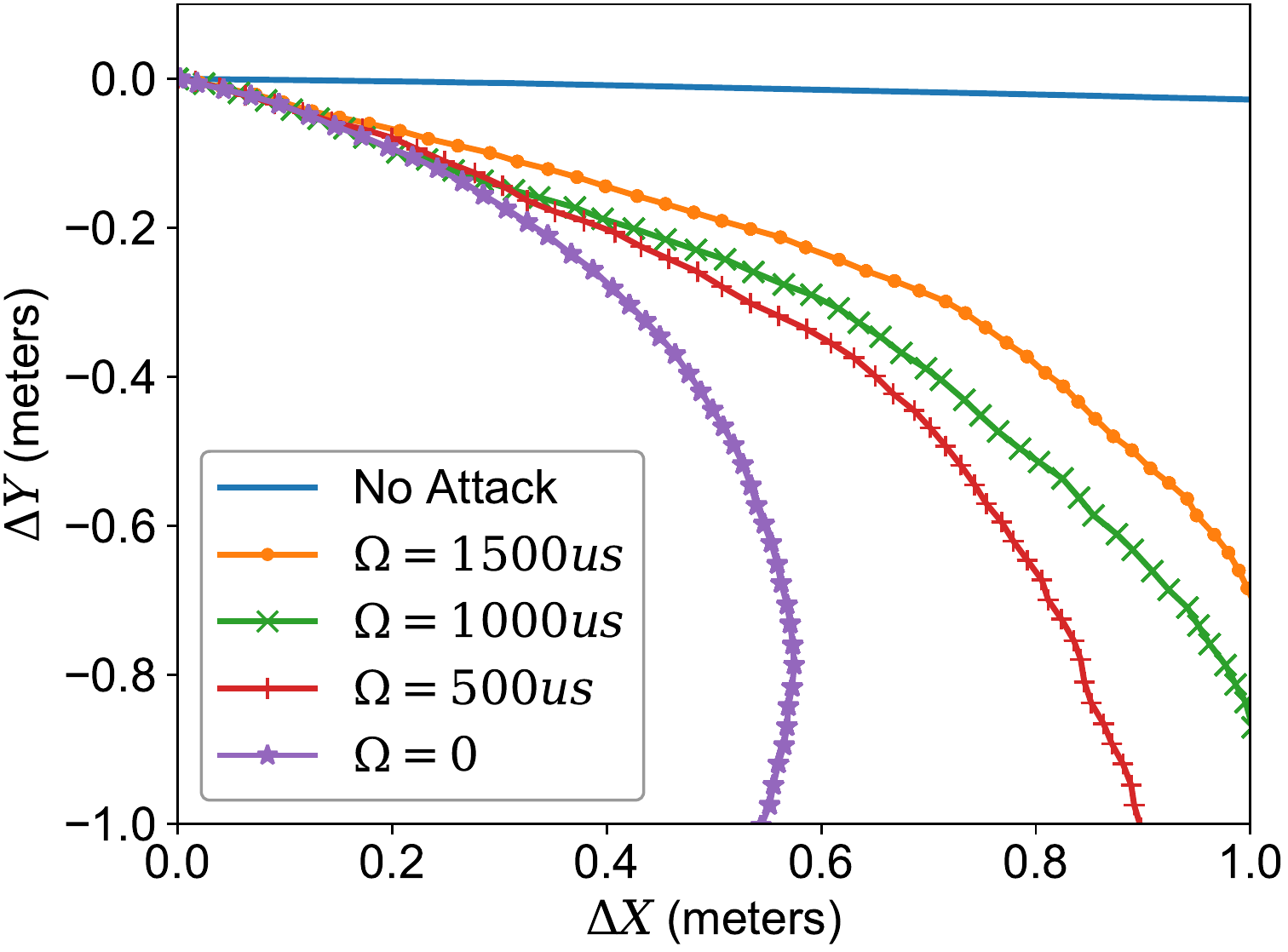}
    \caption{The RC car's trajectories without an attack (the blue line) and with attacks under various $\omega_e$ settings. In this experiment, the car's target is to move straight along the X-axis while the attacker tries to override the steering to make the car turn right.}
    \label{fig:rover_trajectory}
\end{figure}

\noindent\textbf{Defense against control overwrite posterior attack}

In this experiment, we demonstrate the real effect of the proposed defense approach against a real attack on an off-the-shelf RC car with Raspberry Pi 4 and Navio 2 sensor board\footnote{https://navio2.emlid.com/}.
The RoverBot software is utilized as the autopilot. RoverBot\footnote{https://github.com/bo-rc/Rover/blob/master/cpp/RoverBot} is a modularized software stack that runs on Raspberry Pi~4 with a Navio~2 sensor board. 
RoverBot autopilot comprises functionally-separated modules which may run in separate processes, such as Radio input, Localizer, Actuator, \emph{etc}..
Communication among different modules implements a publish-subscribe mechanism using FastDDS\footnote{https://github.com/eProsima/Fast-DDS} framework. 
To perform autonomous waypoint navigation, the Intel RealSense T265 tracking camera\footnote{https://www.intelrealsense.com/tracking-camera-t265/} is connected to the Raspberry Pi 4 computer to provide localization. 
The Intel RealSense SDK 2.0\footnote{https://github.com/IntelRealSense/librealsense} is used to stream the vehicle's real-time poses RoverBot autopilot system, which drives the vehicle to waypoint locations.
 
We launch the control output overwrite attack \cite{chen2018scheduleak} that aims to override the PWM outputs governed by the Actuator task on the car system. 
To create a simpler environment for evaluating the attack and defense results, only the Actuator task is deployed as a SCHED\_FIFO real-time task while others are run as non real-time tasks. 
The Actuator task runs at a frequency of 100Hz and has an average execution time of around 167us. 
The container that runs the Actuator task is configured with rt\_runtime as 400ms, which ensures the task's execution is not throttled. 
To infer the Actuator task's initial offset, we launch a \emph{ScheduLeak} attack as non real-time task in a separate container. 
The obtained initial offset is then used to launch the control output overwrite attack. 
In this attack, the attacker aims to override the steering to make the car turn right while the car is set to move straight. 
The experiment results are shown by the car's trajectories recorded under different test settings as displayed in Figure~\ref{fig:rover_trajectory}. 
The blue line shows the car's trajectory without an attack as a reference.
As the figure shows, the attack can make a sharp right-turn when no protection is involved ($\win=0$).  
As the window length increases, the turn is becoming flat and shaky. This is because the attacker is no longer occupying the $AEW$ and the resulting PWM signal mixes the updates from both the Actuator task and the attacker.
As a result, the attacker is not able to gain full control of the car at will.


\subsection{Simulation}
In this section, we use simulated executions of randomly generated tasksets to showcase how strict enforcement of attack effective window ($AEW$) changes the schedulability of tasksets with \emph{Rate Monotonic} (\emph{RM}) policy. Then we relax the $AEW$ enforcement, allowing all tasks to run within $AEW$, noting if and how long, untrusted tasks execute within $AEW$. This coverage metric is then used to compare \emph{RM} with \emph{Coverage Oriented} (\emph{CO}) Scheduling policy described in~Algorithm~\ref{alg:policy}.

\subsubsection{Schedulability with $AEW$ enforcement}
\label{sec:rm_schedulability}

\begin{figure}[tp]
\centering
\includegraphics[width=\linewidth]{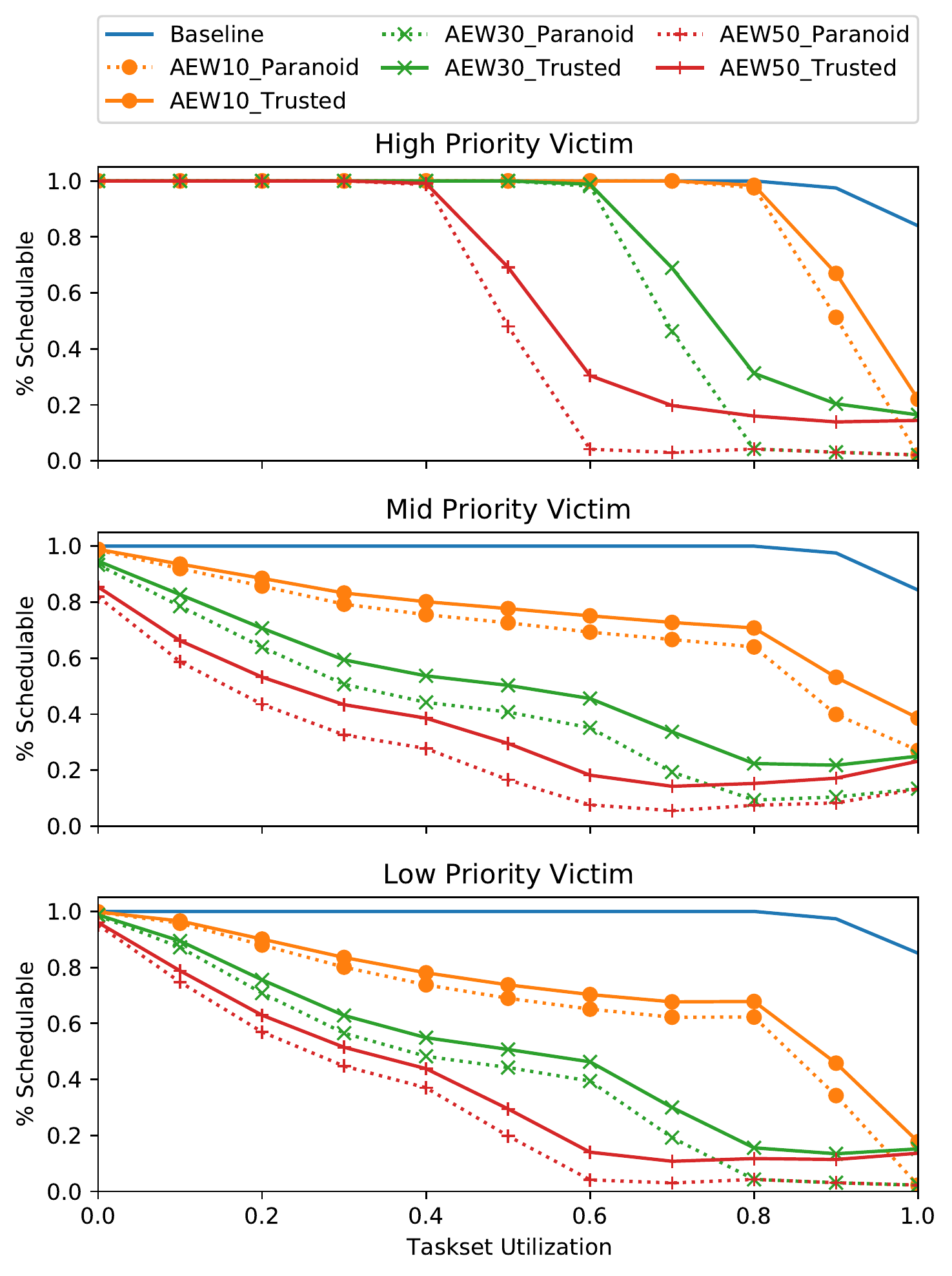}%
\caption{\label{fig:sim_sched} Schedulability of randomly generated tasksets with Rate Monotonic (RM) Scheduling policy.
Baseline refers to standard \emph{RM} scheduling, without any task being protected. For all other cases a victim task is chosen as a High, Medium or Low priority task. $AEW$ sizes as a percentage of victim's period are noted in the legend. In Paranoid scenario no tasks other than the victim are allowed to execute within $AEW$, but in the Trusted scenarios, trusted tasks are allowed to run inside of $AEW$. 20\% of the tasks in a taskset are considered trusted, chosen randomly.}%
\end{figure}

In Figure~\ref{fig:sim_sched}, we use a random task generation software to create 1,000,000 random tasksets for each simulation. 
All tasks are periodic and follow the Liu and Layland task model assumptions~\cite{liu1973scheduling}.
Utilization for each task is selected using \emph{UUniFast} algorithm~\cite{bini2005measuring}.
For each taskset the number of tasks is randomly chosen $\in [2,3,...,10]$.
Periods are also chosen randomly $\in [1,2,...,1000]$ time units with additional constraint to have a hyperperiod of 1000.
Taskset's schedulability is determined by simulating the taskset's execution over the taskset's~\mbox{hyperperiod}.

Figure~\ref{fig:sim_sched} shows the ratio of tasksets (Y-Axis) at each 0.1 utilization interval (X-Axis), which passes the schedulability test.
To show the victim task choice's impact, the victim is chosen as the highest priority, middle priority, and second to lowest priority task in the system. 
Results are plotted for each.
$AEW$ is based on a percentage of the period of the victim task.
$AEW$ percentages $\in [10, 30, 50]$ are explored and noted in the legend \emph{e.g.,} AEW10 implies $AEW$ is $10\%$ of victim task period.
$20\%$ of all tasks are selected as trusted tasks at random, while the victim task itself is always allowed to run within $AEW$.
Baseline refers to \emph{RM} scheduling policy with no victim task or $AEW$ restrictions. Paranoid scenario prohibits the execution of any task other than the victim in $AEW$. In Trusted scenarios, other trusted tasks are also allowed to run within the $AEW$.

\textbf{Discussion of results:}

$AEW$ constraint disallows untrusted tasks to run within the $AEW$ independent of priority.
There are two primary reasons for schedulability changes due to $AEW$.
\textit{First,} the $AEW$ may not be fully utilized by trusted tasks, leading to unusable time in the schedule. When the victim is the highest priority task, this is the only effect observed.
\textit{Second,} $AEW$ also disallows higher priority untrusted tasks to run within it, causing further scheduling failures. 
When the victim is a medium or low priority task, this effect is observed and can cause scheduling failures for low utilization tasksets.
The trusted policy allows trusted tasks to run within the $AEW$ hence the improved schedulability compared to the paranoid policy for the same~setup.


\subsubsection{Coverage Oriented Scheduling Policy}
\label{sec:sim_cover}

We further simulate the \emph{CO} scheduling policy described in Algorithm~\ref{alg:policy} and compare it to \emph{RM}.
In each case, a high priority task is chosen as victim with AEW window sizes a percentage of the task period as before.
The goal of this simulation is to compare for \emph{RM} and \emph{CO} policies, the fraction of the $AEW$ that would be utilized by untrusted tasks when the scheduler does not explicitly protect the $AEW$, rather only records when untrusted tasks are run within the $AEW$
As noted before, \emph{CO} attempts to cover the $AEW$ by executing trusted tasks within $AEW$ but without sacrificing schedulability.



$AEW$ sizes are noted in the legend. The ratio of $AEW$ covered by untrusted tasks (Y-Axis) averaged over tasksets grouped by utilization (X-Axis) are plotted for both \emph{CO} and \emph{RM} for different $AEW$ sizes.

\textbf{Discussion of results:}
\emph{CO} is able to cover more of the $AEW$ with trusted tasks or avoid execution of untrusted tasks within this time by executing them before the victim. Due to the high percentage of untrusted tasks (80\%), the difference is eventually small.
\begin{figure}[tp]
    \centering
    \includegraphics[width=.9\linewidth]{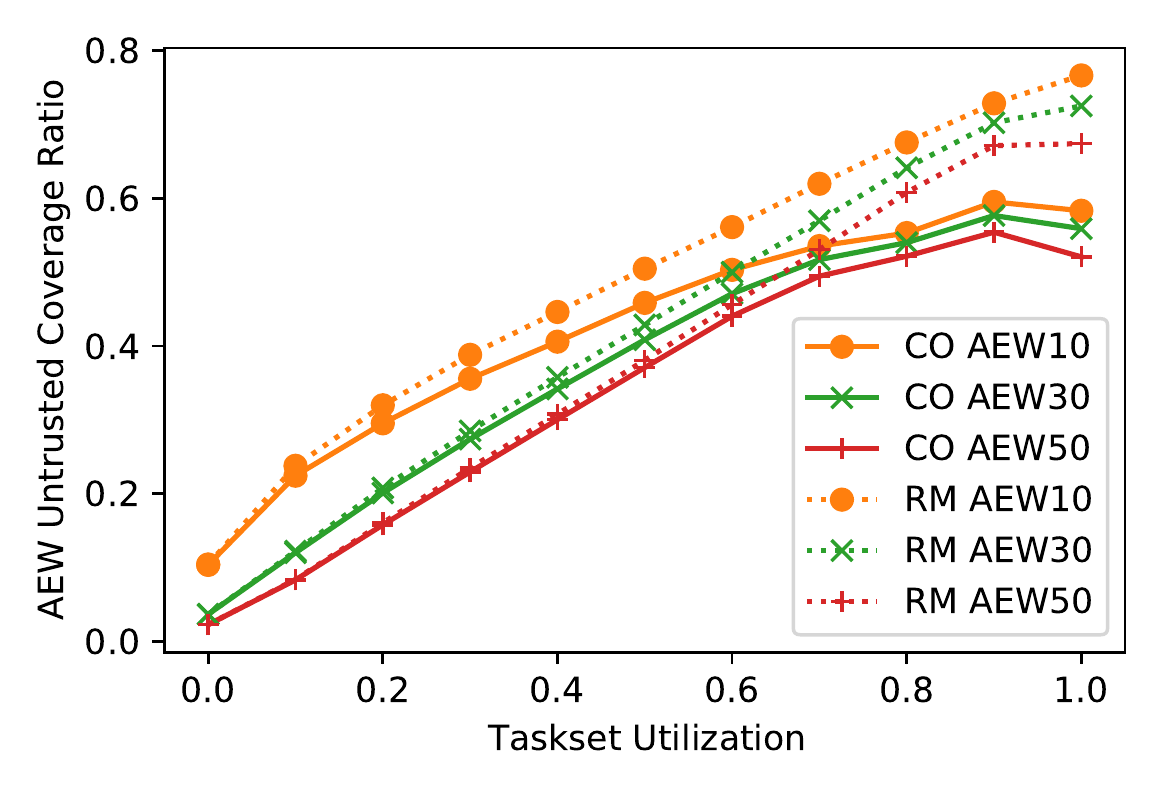}
    \caption{\label{fig:sim_aew_cover}
        Ratio of $AEW$ time covered by Untrusted tasks (Y Axis), averaged for tasksets grouped by utilizations (X Axis). Rate Monotonic (RM) scheduling policy is compared against Coverage Oriented (CO) Scheduling Policy described in Algorithm~\ref{alg:policy}.}%
\end{figure}

\section{Related work}
\label{related}

Side-channel attacks have been considered in the security community as one of the major threats. A variety of them has been studied in the past in~\cite{hu1992reducing, kim2012stealthmem, kocher1996timing}. Solutions such as cache flushing~\cite{hu1992lattice} and hardware/architectural~\cite{mohan2013s3a, suh2004secure, yoon2013securecore, zimmer2010time} modifications have been proposed as a defense mechanism without real-time constraints in mind.

The first work that demonstrated the leakage of information when scheduling tasks in a real-time environment is~\cite{son1998partial}. 
To defend against fixed-priority scheduler from leaking information, authors in~\cite{volp2008avoiding} suggest the use of system idle thread. 
This approach does not consider what happens after the victim task has been completed. 
Similarly, works in~\cite{mohan2014real, pellizzoni2015generalized} suggest defending against the schedule-based information leakage between the high and low security tasks by the introduction of flush tasks. 
This mechanism, however, introduces large overheads, resulting in poor response time of all the tasks in the system and effectively reducing system schedulability. 

Another category of work to defend against the schedule-based attacks is to randomize the schedule~\cite{yoon2016taskshuffler, yoon2019taskshuffler++, chen2018reorder}. 
However, these randomization-based approaches are not very effective and can easily be susceptible to attacks~\cite{nasri2019pitfalls}. Our proposed work does not follow a schedule-randomization-based approach but rather tries to defend against the schedule-based attack by introducing the attack effective window and not allowing the attacker to run during this window. 

From the system's schedulability point of view, some previous works have considered limited preemption~\cite{abdi2017restart, Buttazzo2013}. 
However, to the best of our knowledge, this is the first work that analyzes or considers blocking the window after executing a certain task (victim).

\section{Conclusion}
\label{sec:conclusion}

A new defense mechanism called \emph{SchedGuard} was introduced to defend against the posterior schedule-based attack using Linux containers.
\emph{SchedGuard} prevents untrusted tasks from execution during the specified $AEW$. 
We provided response time analysis for both the paranoid case in which no tasks are allowed to run during $AEW$ and the trusted execution case where only trusted tasks can execute during $AEW$.
We also proposed a novel scheduling policy that provides best-effort protection in the situation where it is not possible to determine the size of $AEW$ while not affecting system schedulability.
We evaluated \emph{SchedGuard} with both simulation and hardware experiments on an embedded platform with real attack.
The results proved the effectiveness of the \emph{SchedGuard} defense mechanism.
In the future, we plan to defend against anterior and pincer attacks using \emph{SchedGuard} and extend it to multicore using gang scheduling \cite{ali2019virtual}.

\section*{Acknowledgment}

The material presented in this paper is based upon work supported by
the Office of Naval Research (ONR) under grant number N00014-17-1-2783 and by
the National Science Foundation (NSF) under grant numbers CNS 1646383, CNS 1932529, CNS 1815891, and SaTC 1718952.
M. Caccamo was also supported by an Alexander von Humboldt Professorship
endowed by the German Federal Ministry of Education and Research. Any
opinions, findings, and conclusions or recommendations expressed in
this publication are those of the authors and do not necessarily
reflect the views of the~sponsors.

\balance

\bibliographystyle{./bibliography/IEEEtran}
\bibliography{ref}

\end{document}